\documentclass[sigplan,nonacm,screen]{acmart}

\usepackage{listings}
\usepackage{booktabs}
\usepackage{tikz}
\usetikzlibrary{arrows.meta,positioning,fit,calc}
\usepackage{xspace}
\usepackage{stmaryrd}
\usepackage{newunicodechar}
\newunicodechar{ε}{\ensuremath{\varepsilon}}
\newunicodechar{Φ}{\ensuremath{\Phi}}
\newunicodechar{σ}{\ensuremath{\sigma}}
\newunicodechar{₀}{\ensuremath{_0}}
\newunicodechar{⊢}{\ensuremath{\vdash}}
\newunicodechar{→}{\ensuremath{\to}}
\newunicodechar{∀}{\ensuremath{\forall}}
\usepackage[capitalise,nameinlink]{cleveref}

\renewcommand\footnotetextcopyrightpermission[1]{}
\setcopyright{none}

\newcommand{\safejit}{SafeJIT\xspace}
\DisableLigatures[-]{encoding=T1,family=tt*}
\newcommand{\code}[1]{\texttt{#1}}
\makeatletter
\newcommand{\lean}{\begingroup\catcode`\_=12 \catcode`\#=12 \catcode`\%=12
  \catcode`\&=12 \catcode`\^=12 \catcode`\~=12 \lean@}
\newcommand{\lean@}[1]{\texttt{#1}\endgroup}
\makeatother
\newcommand{\cc}[1]{\lstinline[style=c]{#1}}
\newcommand{\mlean}[1]{\text{\upshape\ttfamily\detokenize{#1}}}

\newcommand{\wand}{\mathrel{-\!\!*}}
\newcommand{\sep}{\mathrel{*}}
\newcommand{\later}{\mathop{\triangleright}}
\newcommand{\always}{\mathop{\Box}}
\newcommand{\pure}[1]{\ulcorner #1 \urcorner}
\newcommand{\wpre}[2]{\mathrm{WP}\;#1\;\{\!\{#2\}\!\}}

\lstdefinestyle{lean}{
  basicstyle=\ttfamily\footnotesize,
  columns=fullflexible,
  keepspaces=true,
  breaklines=true,
  breakatwhitespace=true,
  breakindent=2em,
  postbreak=\mbox{},
  mathescape=false,
  morekeywords={def,theorem,spec_theorem,spec_holds,structure,inductive,where,fun,let,if,then,else,by,variable,instance,abbrev},
  keywordstyle=\bfseries,
  morecomment=[l]{--},
  commentstyle=\itshape\color{black!60},
  literate=
    {⊢}{{$\vdash$}}1
    {∗}{{$*$}}1
    {-∗}{{$\wand$}}2
    {⌜}{{$\ulcorner$}}1
    {⌝}{{$\urcorner$}}1
    {∀}{{$\forall$}}1
    {∃}{{$\exists$}}1
    {□}{{$\Box$}}1
    {▷}{{$\triangleright$}}1
    {∧}{{$\wedge$}}1
    {∨}{{$\vee$}}1
    {¬}{{$\neg$}}1
    {→}{{$\to$}}1
    {↔}{{$\leftrightarrow$}}1
    {≤}{{$\le$}}1
    {≠}{{$\ne$}}1
    {⟨}{{$\langle$}}1
    {⟩}{{$\rangle$}}1
    {σ₀}{{$\sigma_0$}}2
    {ε}{{$\varepsilon$}}1
    {Φ}{{$\Phi$}}1
    {∈}{{$\in$}}1
    {⦃}{{$\{\!|$}}1
    {⦄}{{$|\!\}$}}1
    {«}{{\guillemotleft}}1
    {»}{{\guillemotright}}1
}

\lstdefinestyle{c}{
  language=C,
  basicstyle=\ttfamily\footnotesize,
  columns=fullflexible,
  keepspaces=true,
  breaklines=true,
  breakatwhitespace=true,
  keywordstyle=\bfseries,
  commentstyle=\itshape\color{black!60},
}

\lstdefinestyle{asm}{
  basicstyle=\ttfamily\scriptsize,
  columns=fixed,
  keepspaces=true,
  morecomment=[l]{\#},
  commentstyle=\itshape\color{black!60},
}

\begin{document}

\title{SafeJIT: A Speculating JIT Engine Verified Down to the Machine Code}

\author{Gabriel Ebner}
\affiliation{%
  \institution{Microsoft Research}
  \country{USA}}
\email{gabrielebner@microsoft.com}

\begin{abstract}
  JIT compilers are among the most attacked components of modern software.
  They compile attacker-chosen programs, speculate on attacker-influenced type
  feedback, and run the result at full native privilege. A single wrongly
  eliminated guard or a missing deoptimization becomes memory corruption.
  We present \safejit, a small JavaScript engine with a speculating baseline JIT,
  together with a machine-checked proof in Lean~4 that its statically linked
  RISC-V and x86-64 executables \emph{never crash, on any input}, in a model
  of a Linux user process. The theorem is about the shipped binaries, executed
  instruction by instruction on Sail ISA models (for RISC-V, the official
  one), and it covers everything in them: the parser and bytecode
  compiler, the interpreter and its type-feedback recording, the JIT compiler,
  all machine code the JIT can generate for any program and any feedback (type
  guards, guard elimination, loop type facts, register allocation with
  deoptimization metadata, inline caches, deoptimization and on-stack
  replacement), a mark-sweep garbage collector, the runtime, and a freestanding
  libc. The C compiler, assembler and linker are not trusted, and neither is
  floating-point arithmetic. The proof uses separation logic (iris-lean) over a
  model of a Linux user process whose stuck states are exactly the ways a
  process can crash, including fetching code that has not been made executable
  or whose instruction cache has not been flushed.

  The engine and its proofs were written by AI coding agents guided by a
  single researcher, and so was this report. An initial RISC-V version took about ten working days; the
  full development, with about 6{,}500 lines of C and 485{,}000 lines of Lean,
  took four weeks. We describe the engine, the theorem and its trusted base, the
  proof architecture, the bugs the proof found, and the methodology. We also
  argue that binary-level verification does not make compiler verification
  obsolete: a JIT is a compiler inside the binary, and most of the proof effort
  goes into showing that the code it generates is safe for every input.
\end{abstract}

\maketitle

\section{Introduction}
\label{sec:intro}

Just-in-time (JIT) compilers exist to run \emph{untrusted} code as fast as
possible: JavaScript and WebAssembly from any web page, and eBPF programs loaded
into the Linux kernel. Their input is chosen by an attacker and their output runs
at full native privilege, outside any type system. A logic error in the
optimizer, such as a guard that was wrongly removed or a type assumption that is
never invalidated, does not cause a wrong answer; it causes memory corruption,
and memory corruption in a browser renderer is the first step of a typical
exploit chain.

The numbers bear this out. All Chrome exploits caught in the wild in 2021--2023
began with renderer memory corruption, and 60\% of those were bugs in V8;
historically about half of the exploited V8 bugs were in one of its
compilers~\cite{v8sandbox}. Microsoft found that about 45\% of the V8 CVEs
since 2019 were JIT-related~\cite{edgesdsm}. Of the 97 zero-days Google tracked
in the wild in 2023, 19 were in browsers and 9 in JavaScript
engines~\cite{tag2024}, and in 2021 six of Chrome's
fourteen zero-days were in V8, including a TurboFan type confusion caused by a
missing deoptimization (CVE-2021-30632)~\cite{p0-2021}. The usual mitigation for
at-risk users is to switch the JIT off, for all sites except user-chosen
exceptions in Apple's Lockdown Mode and in the strict level of Edge's enhanced
security mode, or for unfamiliar sites in its balanced
level~\cite{lockdown,edge-enhanced}, at a cost of 1.5--10$\times$ or more on
compute-heavy code~\cite{v8sandbox}. Memory-safe implementation
languages do not help: ``memory safety cannot be guaranteed by the compiler if a
compiler is directly part of the attack surface''~\cite{v8sandbox}. Fuzzing
\cite{fuzzilli,jitpicking} and sandboxing~\cite{v8sandbox} reduce the damage. A
proof can remove the bug class.

Prior verification work has covered important parts of this problem
(\cref{sec:related}). Jitawa~\cite{jitawa} verified a Lisp runtime and its JIT
down to x86-64 machine code, but compiles before execution, without feedback
or speculation. Icarus~\cite{icarus} verified the type-specialized inline-cache
stubs of SpiderMonkey's CacheIR generators against models of the
macro-assembler. Work on \emph{speculation and deoptimization}, from
sourir~\cite{sourir} to CoreJIT~\cite{corejit,corejit-native}, proves semantic
preservation across speculative tiers; CoreJIT's latest version reasons down to
CompCert's x86 assembly semantics, trusting the step from assembly to
installed bytes and the runtime primitives. Work on JITs for eBPF and
WebAssembly, such as Jitk~\cite{jitk}, Jitterbug~\cite{jitterbug} and
\cite{veriwasm,vwasm,crocus}, reaches machine code and production compilers,
but does not verify speculation with deoptimization. None of these proves a
theorem about the whole \emph{binary} of a speculating engine for a dynamic
language, at the level where an attacker actually operates.

\paragraph{This paper.}
We present \safejit, a small JavaScript engine with a speculating JIT compiler,
and a machine-checked proof in Lean~4~\cite{lean4} that its compiled RISC-V
(RV64) and x86-64 executables never get stuck on any input:

\begin{lstlisting}
def BinarySafe : Prop :=
  ∀ argv fs σ₀,
    Init Image.entry Image.segments argv fs σ₀ →
    ∀ c, Reach (Loop, σ₀) c → PrimStep.NotStuck c

theorem memory_safe : BinarySafe         -- RISC-V
theorem memory_safe_final : BinarySafe   -- x86-64
\end{lstlisting}

\noindent
Here \lean{Image} is the executable, \lean{Init} says that $\sigma_0$ is the
state right after Linux has executed it with arguments \lean{argv} and file
system \lean{fs}, \lean{Reach} is reachability in the operational semantics
of the process, and \lean{NotStuck c} says that configuration \lean{c} has
exited or can take a step (\cref{sec:theorem}).

For every command line (including the debugging flags \code{--trace-jit},
\code{--dump-jit} and \code{--dump-bytecode}) and every file system, the
statically linked ELF executable, run instruction by instruction on a
Sail ISA model~\cite{sail,sail-riscv,sail-x86} (the official one for RISC-V), never executes an illegal
instruction or traps, never touches memory that is unmapped or has the wrong
protection (including fetching from JIT code that has not been made executable),
never fetches an instruction whose bytes were written since the last
instruction-cache flush (on RISC-V, where this is required), and only makes
well-formed system calls. This holds for
any JavaScript program the user supplies and for any type feedback that program
produces. The proofs contain no \lean{sorry} and no axioms beyond Lean's
standard ones and the auxiliary axioms that \lean{bv_decide} introduces for its
verified LRAT checker.

\safejit is a demonstrator: the engine is small. But the theorem has the shape one
would want for a production engine, and it covers the part of a JIT that is
hardest to get right. \safejit records type feedback in its interpreter,
compiles hot functions to native code that \emph{speculates} on the observed
types, eliminates redundant guards using abstract types (within basic blocks and,
on RISC-V, across loop back-edges), allocates locals and a virtual operand stack
to registers, caches property lookups on hidden classes, \emph{deoptimizes} to
the interpreter when a guard fails, and enters compiled code in the middle of a
loop by \emph{on-stack replacement} (OSR). CVE-2021-30632, a missing deopt
after a map change~\cite{p0-2021}, is a bug in exactly these mechanisms.

\paragraph{Contributions.}
\begin{itemize}
\item An end-to-end safety theorem for the binaries of a complete JIT engine for
  a dynamic language, stated against Sail ISA models, for all inputs,
  with neither the C toolchain nor floating point in the trusted base
  (\cref{sec:theorem}). To our knowledge it is the first such theorem for an
  engine that speculates, deoptimizes and performs OSR.
\item A proof architecture that makes this tractable
  (\cref{sec:proof,sec:jit}): a per-instruction bridge from Sail to an abstract
  ISA; a separation logic over whole instructions in which code, including JIT
  code, is a persistent resource; modular specifications that build in
  parallel; GC epochs for a verified collector; and a two-layer JIT proof, in
  which a Lean model of the JIT is proved to emit safe code for every
  well-formed function and every feedback vector, and the binary's JIT is
  proved to write exactly the model's code.
\item Two targets with very different ISAs, RISC-V and x86-64, sharing the
  architecture-neutral parts of the logic (\cref{sec:x86}).
\item Bugs and gaps found during verification (\cref{sec:findings}), among them an
  unchecked instruction-cache flush and unchecked assumptions about the size of
  the address space.
\item An account of how the system was built: almost entirely by AI coding
  agents, with one researcher guiding them, in four weeks
  (\cref{sec:agents}). The proof is about 75 times the size of the C code;
  that would be prohibitive for people, but it is affordable for agents.
\item An argument that binary-level verification, which factors the C compiler
  out of the trusted base, does not make compiler verification obsolete. A JIT
  is a compiler inside the binary. Most of our effort is
  compiler-verification work, done inside a binary-level proof
  (\cref{sec:discussion}).
\end{itemize}

\paragraph{Safety, not correctness.}
The theorem is safety, not functional correctness: we do not prove that
\safejit computes what the ECMAScript specification says. \lean{NotStuck} by
itself rules out executing non-executable or stale bytes, touching unmapped or
wrongly protected memory, and ill-formed system calls. It does not rule out a
store to the wrong object on a writable page, or a well-formed \code{write} of
the wrong data. The proof does establish much more internally: in its
invariants, every value is well tagged and every pointer refers to a live
object of the recorded kind, every memory access is justified by a points-to
fact owned by the accessing code, and the JIT's output is exactly what the
verified model emits (\cref{sec:proof,sec:jit}). These are what rule
out the memory corruption on which JIT exploits are built, but they are
invariants of the proof, not part of the theorem statement.

\paragraph{Artifact.}
We plan to release the engine, the proofs and the tooling as open source;
they are not yet public. C paths in this paper are
relative to the repository root, and Lean paths to \code{proofs/SafeJIT/} (for
RISC-V) or \code{proofs/x86/SafeJITX86/} (for x86-64).

\paragraph{Use of generative AI tools}
Generative AI was used throughout this work. The engine, the proofs and the tooling were written by
AI agents (\cref{sec:agents}). This report was also written by an AI agent,
Claude Opus~5.5 in the GitHub Copilot CLI, from the repository, its history
and the agent session logs, at the author's direction; it was revised in
response to reviews that were themselves produced with AI agents.

\section{Background}
\label{sec:background}

\subsection{Speculating JIT engines}

Production JavaScript engines (V8, SpiderMonkey, JavaScriptCore) execute a
function first in an interpreter. While interpreting, they record
\emph{type feedback}: which types each arithmetic operation, comparison or
property access has seen. Once a function is hot, a JIT compiler translates it
to machine code that \emph{speculates} that the future resembles the past. If
an addition has only ever seen small integers, the compiler emits a 32-bit
integer addition, preceded by \emph{guards} that check the operands' tags and
followed by an overflow check. A guard that fails \emph{deoptimizes}: it rebuilds
the interpreter's frame from the machine state and resumes in the interpreter at
the corresponding bytecode instruction~\cite{deopt92}. Conversely, a function
that is stuck in a long-running loop can switch from the interpreter into
freshly compiled code at a loop header, by \emph{on-stack replacement} (OSR).
Property accesses on objects are accelerated by \emph{hidden classes}
(shapes, maps) and \emph{inline caches}, which remember the shape seen at an
access site and the slot where the property lives~\cite{smalltalk84,self89}.

Every one of these mechanisms has an invariant that is easy to break. A guard
may be removed because an analysis wrongly concludes it is redundant. The
deopt metadata, which maps machine registers back to interpreter slots, may be
wrong at a particular exit. An OSR entry may assume facts that hold on the
normal entry path but not on the one taken from the interpreter. A cached shape
may outlive the assumption that justified it. The resulting code then reads or
writes memory based on a value whose type it has misjudged. That is a type
confusion, and it is the standard first step of a JIT exploit.

\subsection{Sail ISA models}

Sail~\cite{sail} is a language for describing instruction set architectures.
The official RISC-V model~\cite{sail-riscv} is maintained by RISC-V International
and is the golden reference for the ISA; for x86-64 we use an
independent Sail model, written from Intel's manual and hardware
behavior~\cite{sail-x86}. Sail compiles models to theorem provers, including
Lean~4. In the Lean backend we use (a free-monad fork of lean-sail inherited
from xv6iris~\cite{xv6iris}), the model is a Lean program in a free monad
whose effects (memory reads and writes, \code{undefined} values, barriers) are
answered by the caller. A machine step (\code{cycle}) fetches, decodes and
executes one instruction and may raise exceptions, which in our setting are
crashes.

\subsection{Separation logic and Iris}

We reason in Iris~\cite{iris}, a higher-order concurrent separation logic, via
its Lean~4 port iris-lean~\cite{iris-lean}. Iris propositions describe
\emph{resources}: $\ell \mapsto v$ owns the memory at $\ell$,
$P \sep Q$ holds if the resources split into parts satisfying $P$ and $Q$, and
$P \wand Q$ (``magic wand'') turns $P$ into $Q$. Persistent propositions
$\always P$ describe knowledge that can be duplicated freely: in our proofs,
that a word of code is at an address, or that an object of a given kind lives at
a pointer. Iris's weakest precondition $\wpre{e}{\Phi}$ says that $e$ is safe to
execute and that $\Phi$ holds of its result, and \emph{adequacy} turns a closed
proof of a weakest precondition into a statement about the operational
semantics, namely that execution never gets stuck. The later modality
$\later P$ together with L\"ob induction ($(\later P \wand P) \vdash P$) gives
recursion and loops without explicit termination measures.

\subsection{MachCSL and xv6iris}

Kaashoek and Zeldovich verified the xv6 kernel at the RISC-V binary
level~\cite{xv6iris} with MachCSL, a concurrent separation logic over the Sail
RISC-V model, again with AI agents doing most of the proof work. \safejit
builds on their setup: we reuse their Sail configuration, their free-monad
lean-sail fork and a few pieces of MachCSL, such as the boot-register
interpreter. The two projects differ in their object: xv6iris verifies a kernel
whose code is fixed when the proof is done, whereas a large part of
\safejit's executed code does not exist until run time.

\section{The \safejit Engine}
\label{sec:engine}

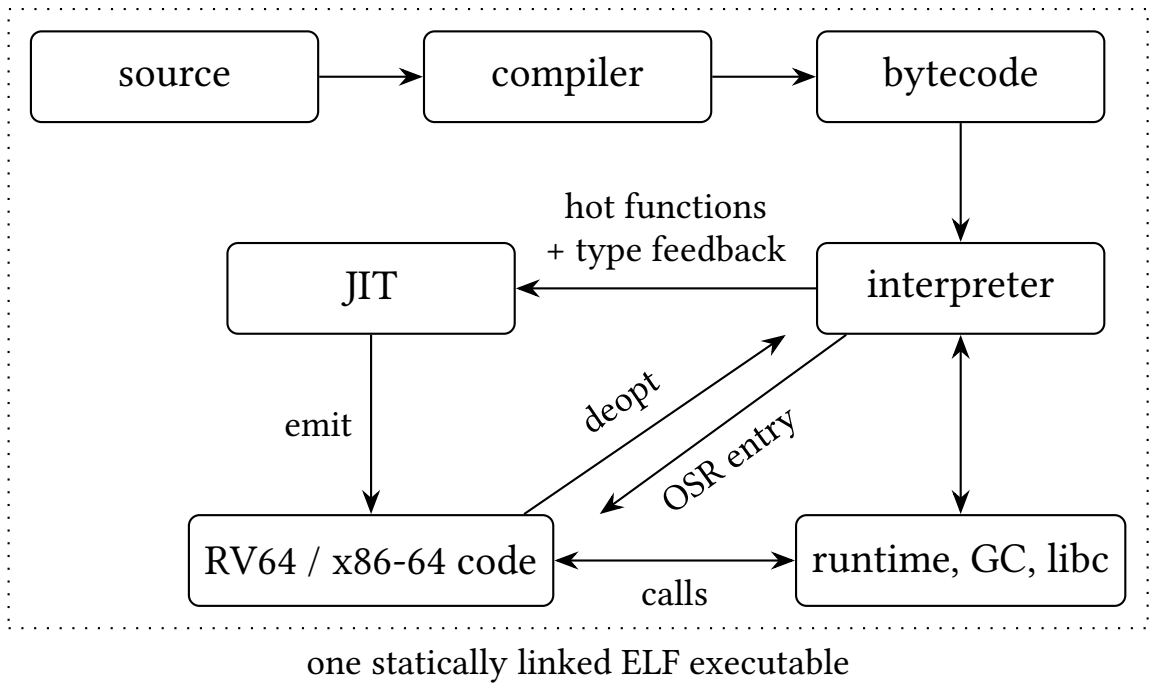
\begin{figure}
\centering
\begin{tikzpicture}[
  box/.style={draw, rounded corners=2pt, minimum height=6mm, minimum width=19mm,
    inner sep=3pt, font=\footnotesize},
  lab/.style={font=\scriptsize},
  >={Stealth[length=1.6mm]}]
\node[box] (src) at (0,0) {source};
\node[box] (cmp) at (2.6,0) {compiler};
\node[box] (bc)  at (5.2,0) {bytecode};
\node[box] (int) at (5.2,-1.4) {interpreter};
\node[box] (jit) at (1.3,-1.4) {JIT};
\node[box] (mc)  at (1.3,-3.2) {RV64 / x86-64 code};
\node[box] (rt)  at (5.2,-3.2) {runtime, GC, libc};
\draw[->] (src) -- (cmp);
\draw[->] (cmp) -- (bc);
\draw[->] (bc) -- (int);
\draw[->] (int) -- node[lab, above, align=center] {hot functions\\+ type feedback} (jit);
\draw[->] (jit) -- node[lab, left] {emit} (mc);
\draw[->] ([xshift=2mm]int.south west) -- node[lab, sloped, below, pos=0.55] {OSR entry}
  ([xshift=3mm]mc.north east);
\draw[->] ([xshift=-2mm]mc.north east) -- node[lab, sloped, above, pos=0.45] {deopt}
  ([xshift=-2mm]int.south west);
\draw[<->] (int) -- (rt);
\draw[<->] (mc) -- node[lab, below] {calls} (rt);
\node[draw, dotted, inner sep=4pt, fit=(src)(cmp)(bc)(int)(jit)(mc)(rt),
      label={[font=\scriptsize]below:one statically linked ELF executable}] {};
\end{tikzpicture}
\caption{Architecture of \safejit. The theorem covers the whole box, including
  the machine code that the JIT generates at run time.}
\label{fig:overview}
\end{figure}

\safejit is a JavaScript engine written in C: about 2{,}500 lines for the
compiler, interpreter, runtime and collector, 1{,}400 for the RISC-V JIT
backend, 1{,}100 for the x86-64 JIT backend, and 1{,}500 for a freestanding C
library. \Cref{fig:overview} shows its structure. A single-pass compiler
translates source code to stack bytecode. An interpreter runs the bytecode and
records type feedback. Hot functions are compiled to native code that speculates
on that feedback, and they return to the interpreter when a speculation fails.
Everything, including the libc, is linked into one static ELF executable that
makes raw system calls. This section describes the engine as an engineer would;
\cref{sec:theorem,sec:proof,sec:jit} describe what is proved about it.

\subsection{Language}

\safejit supports a subset of JavaScript that is large enough to write
realistic benchmarks: \code{let}/\code{const}/\code{var}; numbers (int32 and
double), strings, booleans, \code{null} and \code{undefined}; arrays with
literals, indexing, \code{.length} and \code{push}; objects with literals and
property reads and writes; top-level recursive functions; \code{if},
\code{while}, \code{for}, \code{break}, \code{continue} and \code{return}; and
all of the arithmetic, bitwise, comparison and logical operators, \code{?:},
compound assignment and \code{++}/\code{--}. Builtins include \code{print},
\code{sqrt}, \code{floor}, \code{Array(n)} and \code{clock}. There are no
closures, exceptions or \code{eval}. The test suite compares the engine's output
against Node.js in four tiering configurations: interpreter only, the default
threshold, a threshold of one, and a threshold of two with effectively unlimited
deopts.

Every resource limit is a clean error rather than undefined behavior: syntactic
nesting (1{,}000), bytecode, constants and locals per function (65{,}535 each),
operand-stack depth (30{,}000), call depth (20{,}000), string length
($2^{30}-25$), nested array-to-string conversions, and source size (1~GiB). Out
of memory is a clean exit as well. These limits matter for the proof: every one
of them is an obligation that the proof would otherwise fail to discharge.

\subsection{Values}

Every value is a 64-bit word, NaN-boxed (\cref{tab:nanbox}). A double is stored
as is. Other types live in the upper NaN space and are distinguished by their top
16 bits, so a type check is a shift and a compare. Pointers are 48 bits wide,
so every heap address must lie below $2^{47}$ (\cref{sec:findings} describes
how the proof found that this was not always guaranteed).

\begin{table}
\caption{NaN-boxed value representation.}
\label{tab:nanbox}
\footnotesize
\begin{tabular}{ll}
\toprule
top 16 bits & meaning \\
\midrule
\code{< 0xFFF9} & double (any bit pattern) \\
\code{0xFFF9} & int32 in the low 32 bits \\
\code{0xFFFA} & false = 0, true = 1, null = 2, undefined = 3 \\
\code{0xFFFC} & pointer to a heap object \\
\bottomrule
\end{tabular}
\end{table}

Every bit pattern below \code{0xFFF9}${}\ll 48$ is a valid double, including
the positive canonical NaN \code{0x7FF8\ldots} and x86-64's default NaN
\code{0xFFF8\ldots}. Only negative NaNs with a nonzero payload in bits 48--50
collide with the tags, and a double whose bit pattern reads back as another
type would be a type confusion. Every double result is therefore checked by an
\emph{integer} test: bits at or above \code{0xFFF9}${}\ll 48$ are replaced by
the canonical NaN. The JIT emits this test after every floating-point
instruction; \cc{mk_double} in C also canonicalizes every NaN. RISC-V
arithmetic produces only canonical NaNs, but x86-64 propagates NaN payloads,
so on real hardware the test is needed. In the proof, the ISA models'
floating-point operations remain opaque functions that may return any bits;
the integer test is what makes safety independent of floating-point
semantics.

\subsection{Bytecode}

The compiler is a Pratt parser that emits stack bytecode directly, one function
at a time. While it emits each instruction it also records the operand-stack
depth before it and whether a jump targets it. Nothing re-checks the bytecode
at run time. The interpreter and the JIT both rely on its well-formedness: every
operand is in range, every path reaches an instruction with the same stack depth,
the stack never underflows, and it never exceeds the function's recorded maximum.
In the proof this becomes a predicate \lean{CodeWF} (\cref{sec:proof-compiler})
that the compiler is proved to establish and that the interpreter and the JIT
assume.

\subsection{Interpreter, feedback and tiering}

The interpreter is a dispatch loop over a frame whose \code{slots} array holds
the locals followed by the operand stack. Arithmetic, comparison, conditional
jump and element-access instructions OR a bit into \cc{fn->feedback[pc]}:
\emph{int}, \emph{double} (also set when int operands overflow), \emph{bool},
\emph{array}, or \emph{other}. After 200 calls, or 200 loop back-edges, a
function is compiled. A function call then enters the JIT code at pc~0, and a
back-edge enters it at the loop header by OSR. When JIT code deoptimizes, the
interpreter throws the code away and resumes at the failing instruction, which
now records the feedback that would have prevented the deopt; the function is
recompiled once it is hot again. After $N$ deopts a function stays
interpreted, where $N$ is set by \code{--max-deopts} and defaults to 8.

Calls between functions, including calls from JIT code, go through
\cc{vm_call}. It checks the call depth, sets up the callee's frame on a separate
VM stack, and enters the JIT code if there is any, or the interpreter otherwise.
JIT code and the interpreter are therefore mutually recursive.

\subsection{The JIT compiler}

The RISC-V JIT (\code{src/jit.c}, 1{,}400 lines) emits RV64IMD code into a buffer,
maps it read-write, copies the code in, switches the mapping to read-execute with
\code{mprotect}, and flushes the instruction cache. The flush's result is
checked; if it fails, the code is unmapped and compilation fails.

\paragraph{Shared frame layout.}
JIT code uses the interpreter's frame layout. Because the stack depth at each pc
is static, each operand-stack slot is a fixed offset from the frame base, kept in
a pinned register. Other pinned registers hold the int tag, the boxed
\code{false}, the frame, a table of runtime helpers, and the globals. At a
bytecode boundary where nothing is cached in registers, the memory frame is
exactly what the interpreter would have. Deoptimization then amounts to storing
the pc and stack depth into the frame and returning, and OSR is a jump to the
native address of a loop header.

\paragraph{Speculation.}
For each instruction, a \emph{strategy} is chosen from its feedback
(\cref{tab:strategies}). Int code checks both operands' tags, performs a 32-bit
operation, and checks for overflow and $-0$. Double code checks that the operands
are numbers, converts ints, and uses the FPU. Array code checks the tag, the
object kind and the bounds, and then loads or stores directly. Instructions
without any feedback were never executed by the interpreter. Their code
deoptimizes unconditionally (a ``soft deopt''), so that a branch that was cold
at compile time does not stay generic forever. On the \code{sieve} benchmark,
soft deopts changed the speedup from 1.3$\times$ to about 20$\times$.

\begin{table}
\caption{Strategies chosen from type feedback.}
\label{tab:strategies}
\footnotesize
\begin{tabular}{lll}
\toprule
feedback & strategy & code \\
\midrule
only int & int & tag guards, 32-bit ALU, overflow/$-0$ checks \\
int, double & double & number guards, convert, FPU op, canonicalize \\
only bool & bool & branch on the raw bits \\
only array & array & tag, kind and bounds guards; direct access \\
none & deopt & deoptimize unconditionally \\
other & generic & call a C runtime helper \\
\bottomrule
\end{tabular}
\end{table}

\paragraph{Guards and deopt stubs.}
A failed guard branches to a small out-of-line stub. The stub writes back any
state that is cached in registers, loads the bytecode pc and the stack depth at
which to resume, and jumps to the shared deopt exit. The exit spills the
register-allocated locals, stores pc and depth into the frame, and returns 1 to
\cc{jit_enter}'s caller.

\paragraph{Guard elimination.}
The compiler tracks an \emph{abstract type} for each stack slot and each local
(unknown, int, double, bool, array), and for each stack slot, the local it was
copied from, if any. After \code{i < n} has been checked as int, a later
\code{i + 1} does not check \code{i} again. Results of int operations are known
ints, \code{a.length} is a known int, and so on. A guard is emitted only if the
abstract type does not already imply it.

\paragraph{Loop type facts.}
At a merge point, the compiler normally forgets all abstract types. On RISC-V,
a merge point may instead begin with \emph{checks} of assumed types for up to 16
locals, which deoptimize if a check fails. The code after the checks (the loop
\emph{head}) knows those types. Entries from the interpreter, fallthrough and
forward jumps run the checks. A backward jump whose own abstract state implies
the facts skips them and jumps directly to the head. \cc{jit_compile} infers the
facts with a first code-generation pass, which records the abstract state at
every jump into each merge point, and then compiles a second time with the
intersection of those states. The facts are only hints: any choice is safe,
because the checks enforce them. If the second pass fails, the compiler falls
back to compiling without facts.

\paragraph{Register allocation.}
On RISC-V, the first five locals live in callee-saved registers. Their frame
slots are stale except right after a spill, which happens before every helper
call (helpers may collect garbage or inspect the frame) and at the deopt exit.
Within straight-line code, the top of the operand stack is \emph{virtual}: a
pending slot is held in one of eight pool registers, or it is a copy of a local
or a constant. Pending slots are flushed to the frame before any operation that
may read the frame or leave straight-line code. The deopt stub of an
instruction materializes exactly the pending slots that exist at that
instruction's start. This is the deopt metadata that production JITs keep, in
its simplest form. A comparison followed by a conditional jump is fused into a
single native branch.

\paragraph{Objects and inline caches.}
Objects have hidden classes: a shape is a node in a transition tree, and each
child adds one property. Shapes are immortal, so a cached shape stays valid
forever. Every property access site has a monomorphic inline cache holding a
shape and a slot (for a store that adds a property, also the shape after the
addition). The caches are \emph{data-driven}: the JIT embeds the address of the
site's cache in the code, and the code reads the cache at run time. It checks
that the receiver is an object whose shape equals the cached one, then loads
or stores the cached slot directly; on a miss it calls the runtime's property
helper, which updates the cache. The code is never patched, and the cache is
shared between the interpreter and JIT code.

\paragraph{An example.}
\Cref{fig:dump} shows the code that \code{--dump-jit} prints for part of the
loop \code{for (let i = 0; i < a.length; i++) s = s + a[i];}. Locals \code{a} and
\code{i} are in registers \code{s0} and \code{s9}. \code{a.length} checks that
\code{a} is an array, and its result goes to a pool register (\code{a5}) instead
of the frame. The comparison does not check \code{i}'s tag, because the loop
fact says that \code{i} is an int, and it is fused with the loop's exit branch.
Element access does not check that \code{a} is an array, because it was checked
in the same block. The deopt stubs at \code{0x150} and \code{0x1b8}
write the pending state back to the frame before they leave.

\begin{figure}
\begin{lstlisting}[style=asm]
 ; 18 GET_LENGTH  feedback={arr} -> array
 114: mv    a0, s0          # local 0 (a) is in s0
 118: srli  t0, a0, 48      # tag
 11c: addi  t1, s4, 3       # s4 = 0xFFF9; +3 = object tag
 120: bne   t0, t1, 150     # guard: object
 124: slli  a0, a0, 16      # unbox pointer
 128: srli  a0, a0, 16
 12c: lwu   t0, 0(a0)       # object kind
 130: li    t1, 1
 134: bne   t0, t1, 150     # guard: array
 138: lwu   a3, 4(a0)       # length
 13c: ...                   # box as int32
 148: mv    a5, a3          # pending stack slot in a5
 14c: j     164
 150: sd    s9, 24(s1)      # deopt stub: materialize
 154: sd    s0, 32(s1)      #   the pending slots,
 158: li    t0, 18          #   pc = 18,
 15c: li    t1, 2           #   depth = 2,
 160: j     78              #   to the deopt exit
 ; 19 LT  feedback={int} -> int     (i known int)
 164: mv    a0, s9
 168: mv    a1, a5
 16c: sext.w a0, a0
 170: sext.w a1, a1
 174: blt   a0, a1, 17c     # fused with JMP_IF_FALSE
 178: j     260             # loop exit
 ; 23..29 GET_LOCAL i; GET_LOCAL a; GET_LOCAL i  (virtual)
 ; 32 GET_ELEM  feedback={arr} -> array  (a known array)
 17c: mv    a0, s0
 180: mv    a1, s9
 ...  bounds check, load a[i] into a5 ...
 1b8: sd    s8, 24(s1)      # deopt stub of GET_ELEM
 ...
\end{lstlisting}
\caption{Annotated \code{--dump-jit} output (RISC-V) for the loop of
  \code{sum(a)}. Comments are ours.}
\label{fig:dump}
\end{figure}

\subsection{Runtime, garbage collector and libc}

The runtime implements the generic semantics: conversions, \code{+} on strings,
comparisons, number printing, arrays, objects and the builtins. Strings, arrays
and objects are collected by a non-moving mark-sweep collector
(\code{src/gc.c}, about 100 lines). It runs only at a safepoint in
\cc{vm_call}, once enough objects have been allocated since the last
collection. At that point every live value is the callee, a slot on the VM
stack, or a global; NaN-boxing makes pointers precise, and no C local, machine
register or JIT code holds a collectible object across a call. Functions,
natives and shapes are immortal. JIT code is never freed: each function is
compiled at most $N$ times, so the code leaked by deopts is bounded by $N$
times a constant multiple of the program size.

The libc (\code{src/libc}) is written to be easy to verify: straight-line code,
bounded loops, fixed-size buffers and no heap use in the number conversions.
\code{\_start} switches to a static 64~MiB stack. \cc{malloc} is a size-class
allocator over 64~MiB \code{mmap}ed arenas with a real \cc{free}. \cc{strtod} and
\cc{printf("\%e")} are exact (correctly rounded), using a fixed 4096-bit bignum.
The C compiler is GCC with \code{-march=rv64g}: no compressed instructions and
no libgcc, so the binary and the JIT's output are both made only of 32-bit
instruction words.

\subsection{The x86-64 backend}

\code{src/x86\_64/jit.c} is a port of the RISC-V JIT to x86-64. It has the same
frame layout, strategies, guards, guard elimination, deopt protocol and OSR, but
neither register allocation nor loop facts. The port has to deal with x86's
peculiarities. \code{idiv} traps on division by zero and on
\code{INT\_MIN / -1}, so the divisor is checked first and the division is done
in 64 bits. Overflow is detected with the OF flag. \code{ucomisd} sets
ZF/PF/CF for unordered operands, so comparisons are chosen so that NaN operands
give false. SSE arithmetic propagates NaN payloads, so the check after every
FP operation is required on real hardware, not just defensive. The stack must be 16-byte
aligned at calls into C helpers. Everything else (compiler, interpreter,
runtime, collector and most of the libc) is the same C code, compiled with
GCC for x86-64-v1.

\section{The Theorem}
\label{sec:theorem}

This section states what is proved and what is trusted. We describe the RISC-V
development; \cref{sec:x86} describes how the x86-64 one differs.

\subsection{A process language in which stuck means crashed}

The theorem is stated over a small-step semantics of a Linux user process
(\code{Proc/Lang.lean}). A process state $\sigma$ consists of the Sail register
file, a byte-addressed memory with per-page read/write/execute protection, and
an abstract kernel state (open files with their offsets and contents, and the
file system). A step either executes one Sail \code{cycle}, or, when the
instruction at the program counter is \code{ecall}, executes a system call
atomically according to a contract. The language is constructed so that every
way a real process can crash is a state with \emph{no} step
(\cref{tab:stuck}).

\begin{table}
\caption{Crashes and how the process model represents them.}
\label{tab:stuck}
\footnotesize
\begin{tabular}{p{0.42\columnwidth}p{0.18\columnwidth}p{0.25\columnwidth}}
\toprule
event & Linux & model \\
\midrule
trap out of U-mode (illegal or misaligned instruction, \code{ebreak}, \ldots) & fatal signal & no step \\
access outside mapped memory or against its protection & \code{SIGSEGV} & no step \\
executing JIT code not made executable & \code{SIGSEGV} & no step (fetch needs X) \\
fetching a byte written since the last I-cache flush & stale code & no step \\
a failed Sail assertion & --- & no step \\
syscall outside the contract (unknown number, bad pointer, bad flags) & \code{-EFAULT}, \ldots & no step \\
\code{exit\_group(n)} & exit & final value $n$ \\
\bottomrule
\end{tabular}
\end{table}

\paragraph{Instruction-cache coherence.}
The Sail RISC-V model executes \code{fence.i} as a no-op and leaves
instruction-cache coherence to the memory model. On SMP Linux \code{fence.i}
is not enough anyway, because the thread may migrate to another hart; user
programs must call the \code{riscv\_flush\_icache} system call. The process
memory therefore tracks \emph{dirty} bytes. Every store marks the bytes it
writes, and an instruction fetch of a dirty byte has no step. Only a
\code{riscv\_flush\_icache} that \emph{returns 0} clears the dirty bits, and the
initial image is clean. A JIT that writes code and runs it without a successful
flush is therefore stuck in the model, as it may be on real hardware.

\paragraph{Address space.}
Linux on Sv57 can hand out addresses up to $2^{56}$: \code{mmap} falls back above
$2^{47}$ once the low window is full, and before Linux~6.6 the stack sat just
below the top of the address space. The model therefore lets the kernel place
the stack and \code{mmap} results anywhere below $2^{56}$, except for the
platform's CLINT region, which the Sail configuration requires to be unmapped.
The program itself must not assume more than it checks.

\subsection{The statement}

Let \lean{Init entry segs argv fs σ₀} say that $\sigma_0$ is a state in which the
kernel has just executed the ELF image with entry point \lean{entry} and loadable
segments \lean{segs}, with argument vector \lean{argv} and file system
\lean{fs}. Safety says that every reachable configuration can step or has
exited:

\begin{lstlisting}
def Safe (entry : Addr) (segs : List Segment) : Prop :=
  ∀ argv fs σ₀, Init entry segs argv fs σ₀ →
    ∀ c, Reach (Loop, σ₀) c → PrimStep.NotStuck c

def BinarySafe := Safe Image.entry Image.segments
theorem memory_safe : BinarySafe

def ElfSafe := Elf.Wf file ∧ Safe (Elf.entry file) (Elf.segments file)
theorem memory_safe_elf : ElfSafe
\end{lstlisting}

\noindent
\lean{argv} ranges over all lists of byte strings, so the theorem covers every
JavaScript program (the program is a file named on the command line), every
flag (including \code{--jit-threshold=N}, \code{--max-deopts=N}, and the
debugging flags that disassemble JIT code), and every file system state.
\lean{memory_safe_elf} restates the theorem over \lean{file}, the raw bytes of the
\code{safejit} executable, read through a small formal ELF loader
(\code{Proc/Elf.lean}). The image that the proofs work with, a list of decoded
instructions in \code{Image.lean}, is generated by an untrusted script;
\lean{decide +kernel} checks that it agrees with what the loader reads from the
raw bytes.

The theorem quantifies over every state after a successful modeled exec, so
it is only as meaningful as \lean{Init} is satisfiable. We check part of
this: \lean{initRegs_nonempty} shows that the register-file part of
\lean{Init} has a witness for every entry point and stack pointer,
established by running the Sail model's own
\code{initialize\_\allowbreak registers} followed by the writes that Linux
performs at exec. We have not built a witness for the full predicate on the
actual image (with loaded segments, a valid stack and an admissible argument
vector); \lean{Init} is written to be a direct transcription of what Linux
does, and arguments are C strings, so they cannot contain NUL bytes.

A safety theorem is also satisfied by an engine that does nothing
interesting, for example a JIT that always fails or always deoptimizes. Two
things rule this out for \safejit. The binary's compiler is proved to write
exactly the code of the executable model \lean{Jit.compile}
(\cref{sec:jit}), and that model is tested against the logged compilations of
the real engine; and the benchmarks (\cref{sec:eval}) show that the JIT code
runs and is 2--25$\times$ faster than the interpreter.

\subsection{The trusted base}

The trusted base consists of the following.
\begin{itemize}
\item \textbf{The Sail RISC-V model} with the F and D extensions, compiled to
  Lean by Sail's Lean backend, and the free-monad Sail library. Our
  regeneration script also fixes a bug in that backend: \code{flen = flen\_bytes
  * 8} was printed without parentheses, so every FP register was eight bits wide.
\item \textbf{The process abstraction}: one U-mode hart with \code{satp = Bare},
  sequentially consistent memory with per-page protection, no signals, no
  interrupts, and the dirty-byte rule for instruction fetches.
\item \textbf{The syscall contract} (\code{Proc/Syscall.lean}) for twelve
  system calls: \code{read}, \code{write}, \code{openat}, \code{close},
  \code{lseek}, \code{mmap} (anonymous only), \code{mprotect}, \code{munmap},
  \code{clock\_gettime}, \code{riscv\_flush\_icache}, \code{exit} and
  \code{exit\_group}. Every call except the two exits may fail with any error code, so the
  program must handle failure everywhere. A successful \code{mmap} returns fresh
  pages anywhere below $2^{56}$. A failed flush leaves the instruction cache
  stale.
\item \textbf{The exec state}: the registers that Linux sets up and the
  System~V initial stack, which can lie anywhere in $[2^{37}, 2^{56})$.
\item \textbf{The ELF loader} (\code{Proc/Elf.lean}) and its well-formedness
  conditions (a RISC-V executable without an interpreter, segments inside the
  file and disjoint, no wrap-around, page-congruent offsets), and the raw file
  dump \code{ElfFile.lean}, which a separate Lean program checks byte for byte
  against the executable.
\item \textbf{Lean's kernel}, and the compiled LRAT checker of
  \lean{bv_decide}, whose uses appear as auxiliary axioms. No proof uses
  \lean{native_decide}.
\end{itemize}

\noindent
The trusted base does \emph{not} include the C compiler, assembler or linker,
since the proof is about their output, or the libc, which is part of the
binary. Nor does it include floating point: the model's softfloat operations
are opaque functions, and no safety argument depends on their results. NaN-boxing
canonicalizes every double with an integer test, and any value derived from a
floating-point result, such as an array index from \code{fcvt}, is
bounds-checked before it is used. The build is reproducible: compiling the
sources produces exactly the bytes in \code{ElfFile.lean}.

\paragraph{What the theorem does not say.}
The theorem does not say that \safejit implements JavaScript correctly, that it
terminates, or that it does not leak memory. Out-of-memory, nesting limits and
JavaScript errors are clean exits, and so are consistent with safety. A
program that loops forever is safe. These are deliberate choices. Safety is the
property that attackers need to violate, and it can be stated without a formal
semantics of JavaScript. Each of these stronger properties is an extension,
not a replacement (\cref{sec:discussion}).

\section{Proof Architecture}
\label{sec:proof}

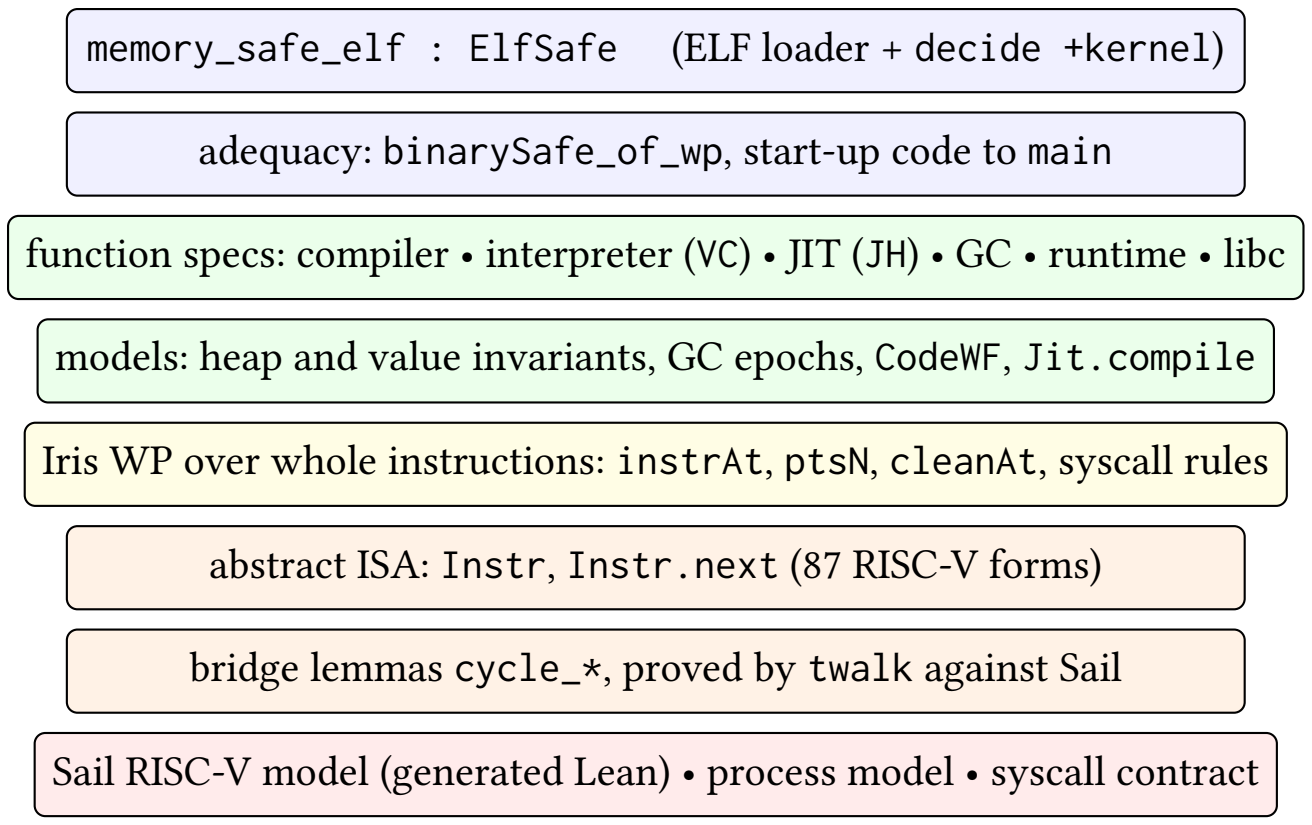
\begin{figure}
\centering
\begin{tikzpicture}[
  layer/.style={draw, rounded corners=2pt, minimum width=0.92\columnwidth,
    minimum height=5.5mm, font=\footnotesize, align=center, fill=#1},
  layer/.default=white, node distance=1.2mm]
\node[layer=blue!6] (top) {\lean{memory_safe_elf : ElfSafe} \quad (ELF loader + \lean{decide +kernel})};
\node[layer=blue!6, below=of top] (ad) {adequacy: \lean{binarySafe_of_wp}, start-up code to \code{main}};
\node[layer=green!8, below=of ad] (fn) {function specs: compiler \textbullet{} interpreter (\lean{VC}) \textbullet{} JIT (\lean{JH}) \textbullet{} GC \textbullet{} runtime \textbullet{} libc};
\node[layer=green!8, below=of fn] (inv) {models: heap and value invariants, GC epochs, \lean{CodeWF}, \lean{Jit.compile}};
\node[layer=yellow!12, below=of inv] (sl) {Iris WP over whole instructions: \lean{instrAt}, \lean{ptsN}, \lean{cleanAt}, syscall rules};
\node[layer=orange!10, below=of sl] (isa) {abstract ISA: \lean{Instr}, \lean{Instr.next} (87 RISC-V forms)};
\node[layer=orange!10, below=of isa] (br) {bridge lemmas \lean{cycle_*}, proved by \lean{twalk} against Sail};
\node[layer=red!8, below=of br] (sail) {Sail RISC-V model (generated Lean) \textbullet{} process model \textbullet{} syscall contract};
\end{tikzpicture}
\caption{The layers of the RISC-V proof. Red is trusted, orange is the
ISA abstraction, yellow is the program logic, green are the program proofs,
and blue is the statement.}
\label{fig:layers}
\end{figure}

\Cref{fig:layers} shows the structure of the proof. At the bottom, the Sail
model executes one instruction as a long program of register reads, memory
events and bit-vector operations. One \emph{bridge lemma} per instruction form
turns this program into an abstract one-step semantics. An Iris program logic
over whole instructions sits on top of that, and the separation-logic
specifications of the binary's functions are proved in that logic.
Adequacy of the logic then gives \lean{Safe}. This section goes through the
layers from the bottom up. \Cref{sec:jit} covers the JIT.

\subsection{From Sail to an abstract ISA}

The Sail model's \code{cycle} function, compiled to Lean, is a program in a
free monad whose effects are register reads and writes, memory events and
nondeterministic choices. To reason about it we use a symbolic executor,
\lean{twalk}, a Lean tactic that proves that a cycle run from a symbolic
register file satisfying some hypotheses ends in a given postcondition for
every answer to every effect. It executes one effect at a time. It folds closed
subterms by kernel reduction and decides branch conditions with
\lean{decide}, then \lean{bv_decide}, then \lean{bv_omega}, using only the
hypotheses that mention the condition's variables. When a condition cannot be
decided it splits on it. Decoding a symbolic instruction word with a symbolic
register field needs special simplification procedures, which select
instruction fields by width.

Each \emph{bridge lemma} proves one instruction form for all operands. For
example, \lean{cycle_addi} covers \code{addi rd, rs1, imm} for all
\code{rd}, \code{rs1} and \code{imm}. It takes 487 steps and about 25~seconds.
The lemma states that the cycle fetches the instruction, preserves the
configuration invariant \lean{Cfg} (U-mode, no translation, the PMP and PMA
settings, interrupts disabled), sets \code{PC} to $\mathit{pc}+4$ and updates
the register file. Loads and stores additionally require a naturally aligned
address in main memory and record the data access. The 87 bridge lemmas cover
every instruction form in the binary and every form the JIT emits: 30 integer
ALU forms, eight control-transfer forms, eleven loads and stores, thirteen
RV64M forms, 24 F/D forms and \code{fence.i}. For the multiplication and
division instructions, a separate file proves that Sail's \code{Int}
arithmetic (truncating division with an overflow clamp) equals the
corresponding \code{BitVec} operation. Floating-point results are the
model's opaque softfloat functions.

A script reads the lemma statements and generates \code{Proc/ISA.lean}: an
inductive type \lean{Instr} with one constructor per lemma, its encoding, a
memory precondition \lean{Instr.Pre} and a functional next-state function
\lean{Instr.next}. A generic lemma \lean{step_instr} then says that, if the
program counter points at executable bytes encoding \lean{i} and
\lean{i.Pre} holds, every cycle ends in the state given by \lean{i.next}. Above
this layer, \emph{no proof mentions Sail}. Every word of \code{.text} is
decoded into an \lean{Instr} (or \code{ecall}) by a script, and the decoding is
checked: every word must decode uniquely and re-encode exactly.

\subsection{An instruction-level separation logic}

We instantiate iris-lean's language interface with the process: an expression
is \lean{AExpr.run r}, a register file about to execute, and a step is one
instruction or one system call. The state interpretation is an authoritative
ghost map over the bytes of memory, with each byte's protection, a ghost copy
of the kernel state (\lean{osIs}), and a ghost set of \emph{clean} bytes.
The basic resources are as follows.
\begin{itemize}
\item \lean{ptsN a dq p n w}: the $n$ bytes at $a$ hold $w$, with fraction or
  discarded fraction \lean{dq}, and the page protection includes \lean{p}.
  \lean{bytesAt}, \lean{slots} and \lean{cstrQ} are byte strings, stack words
  with arbitrary contents, and NUL-terminated strings.
\item \lean{cleanAt a}: byte $a$ is read-only and clean, so it can be fetched. It
  is persistent, and only the boot rule and the rule for a \emph{successful}
  \code{riscv\_flush\_icache} produce it.
\item \lean{instrAt pc w}: a persistent, executable, clean, read-only word $w$
  at $\mathit{pc}$. The same rules therefore fetch from \code{.text} and from
  JIT code.
\end{itemize}

\noindent
There is one rule per class of instructions. For example, a load requires
\lean{instrAt r.pc i.encode} and a points-to for the address
\lean{i.access r} computes, and continues under a later with the next state:
\[
\begin{array}{l}
\mlean{instrAt}~\mathit{pc}~(\mathit{enc}~i) \wand \mlean{ptsN}~a~q~p~n~w \wand {} \\
\quad \later\bigl(\mlean{ptsN}~a~q~p~n~w \wand \wpre{\mathit{next}(i, r, w)}{\Phi}\bigr) \\
\quad\wand \wpre{r}{\Phi}
\qquad\text{if } \mathit{access}(i, r) = \mathit{load}~a~n.
\end{array}
\]
Syscall rules follow the contract, and each covers the failure case as well as
success. For example, \lean{wp_sys_mmap} hands out fresh zero-filled pages
below $2^{56}$ on success. The adequacy theorem, together with a proof that the
start-up code \code{\_start} and the prologue of \code{\_\_libc\_start\_main}
reach \code{main} with the expected resources, reduces \lean{BinarySafe} to a
weakest precondition at \code{main}.

\paragraph{Function specifications.}
A function specification is a weakest precondition at the function's entry
address, in continuation-passing style. Here is \code{strlen}:
\begin{lstlisting}
spec_theorem strlen_spec (x f : Nat → BitVec 64) (dq : DFrac) (p : Perm)
    (hpr : p.r = true) (s : List Byte) (hx0 : x 0 = 0) (Φ : Byte → IP) :
  ⊢ textCode -∗ cstrQ (x 10) dq p s -∗
    (cstrQ (x 10) dq p s -∗ ∀ x', ⌜Clobbers x x' [10, 14, 15] ∧
        x' 10 = BitVec.ofNat 64 s.length⌝ -∗
      WP (AExpr.run ⟨x 1 &&& ~~~1#64, x', f⟩) {{ Φ }}) -∗
    WP (AExpr.run ⟨&strlen#64, x, f⟩) {{ Φ }}
\end{lstlisting}
\noindent
\lean{textCode} is persistent ownership of all of \code{.text}. The continuation
is the weakest precondition at the return address in \code{ra} (\lean{x 1}),
with the new register file \lean{x'}, which may differ from \lean{x} only in the
listed registers. A spec thus says precisely which registers a function may
clobber. This is the ABI as the compiler actually used it, which can be narrower
than the standard one. \lean{&strlen#64} is a \emph{symbolic address}: a
notation, generated from the ELF symbol table, that expands to the address of
\code{strlen} in this build. Proofs refer to code addresses only in this
form, such as \lean{&fmod+0xdc} or \lean{&"(null)"#64}. After a relink most
proofs still check, unless the code inside a function moved.

\paragraph{Executing code.}
Tactics run straight-line code: \lean{sl_step} executes one instruction,
finding the right points-to in the context by the address the instruction
accesses, and \lean{sl_steps} runs up to the next branch, jump or
\code{ecall}. Stack frames are read off the code rather than typed into the
proofs: \lean{frameSaves &fn} scans \code{fn}'s prologue for its register
spills, so a proof does not hard-code the frame layout that the compiler
chose. Loops are proved by induction or by Löb induction.

\paragraph{Modular specifications.}
With hundreds of functions and specifications, the proofs must be checked in
parallel, and a change to one function's proof must not trigger a rebuild of
its callers' proofs. We built a small Lean extension for this
(\code{Holds.lean}). An interface module states a specification with
\lean{spec_theorem}, as above but without a proof. This defines its
statement \lean{foo_spec.Stmt} and a theorem \lean{foo_spec} that takes
\lean{[Holds foo_spec.Stmt]} as an instance argument. A proof module imports
only the interfaces of its callees, assumes their statements as section
variables, and proves its own with \lean{spec_holds foo_spec := by ...}, which
registers an instance of \lean{Holds} that is parametric in the callees'
\lean{Holds} assumptions. The top-level module imports all proof modules and
closes the specifications callees first with \lean{spec_link}, after which
instance resolution assembles a closed proof term. No axioms are involved:
a missing proof leaves an instance that cannot be resolved, and the main
theorem does not typecheck. The RISC-V proof has 376 \lean{spec_holds}
declarations and the x86 proof has 222. Besides build parallelism, this split
was what made it possible for several AI agents to work on the proof at once
(\cref{sec:agents}).

\subsection{The runtime: heap, values and garbage collection}

\paragraph{Allocator.}
The \code{malloc} specification exposes the size-class design: the invariant
\lean{allocInv} owns the current arena and the 95 free lists, and
\code{malloc(n)} returns either 0 or an \lean{mblock}, a zero-filled,
16-byte-aligned block behind a header, owning its padding up to the class
capacity. \code{free} takes an \lean{mblock} back. Clients that free memory,
such as the JIT's scratch buffers and \code{realloc}, keep the \lean{mblock}
rather than forgetting it.

\paragraph{Heap and values.}
The heap invariant \lean{heapInv} owns an authoritative ghost map from
addresses to object kinds (string, array, object, function, native). Every live
object has a persistent fragment:
\begin{lstlisting}
def objAt (p : Addr) (k : Kind) : IP :=
  ⌜p.toNat % 16 = 0 ∧ 16 ≤ p.toNat ∧ p.toNat < 2 ^ 47⌝ ∗ tyIs p k ∗ objRo p k
def valOk (v : BitVec 64) : IP :=
  ⌜IsNum v⌝ ∨ ⌜IsSpecial v⌝ ∨ ∃ k, ⌜v >>> 48 = 0xFFFC#64⌝ ∗ objAt (v &&& ptrMask) k
\end{lstlisting}
\noindent
\lean{objRo} gives fractional, read-only access to an object's immutable part:
a string's bytes, or a function's bytecode, constants and metadata. The
mutable parts (an array's length, capacity and element block, an object's
shape and slots, a function's feedback bytes, counters and \code{fn->jit}
pointer) live in the heap invariant and are borrowed from it. For a function,
the read-only part includes the facts that its bytecode is well formed
(\lean{CodeWF}, below), that the jump-target table is consistent with it,
and that every property-access opcode names an existing inline cache.
\lean{valOk} is the value invariant: the VM stack, globals, array elements,
object slots and return values all hold valid values.
Its pointer case bounds the address below $2^{47}$, which is where the proof
uncovered the address-space bug (\cref{sec:findings}).

Inline caches are cells \lean{icCell a n} that hold either nothing or a
shape and slot index for property \lean{n}, or a transition to a shape that
adds \lean{n}. Shapes are immortal and persistent (\lean{shapeAt sh ns}), and
the cell's invariant \lean{icOk} says that a cached slot index is below the
number of properties of the cached shape, so an object with that shape has the
slot. JIT code reads the cache at run time, so this invariant is what makes
an inline load from a cached slot safe, whatever the cache holds at that moment.
When the runtime's property helper updates a cache, a ghost map of
\emph{loans} records that the cell is lent out of its function's body.

\paragraph{GC epochs.}
\lean{objAt} is persistent, but objects are freed by the collector. We
handle this with \emph{epochs}. The heap ghost map is not a global constant:
\lean{objAt}, \lean{valOk} and every invariant mentioning them are
parameterized by an instance \lean{HG : HeapGS}, the current epoch's ghost name.
A garbage collection allocates a fresh ghost map that contains only the marked
objects. Persistent facts about the old epoch become useless, except for those
that are transferred. The transfer is
\begin{lstlisting}
def epT (HG HG' : HeapGS) (R : List (BitVec 64)) : IP :=
  □ ∀ p k, tyIs (HG := HG) p k -∗
    (⌜k.Imm ∨ RootP R p⌝ ∨ ∃ q fi, ⌜RootP fi.consts p⌝ ∗ tyIs (HG := HG) q (.fn fi)) -∗
    tyIs (HG := HG') p k
\end{lstlisting}
\noindent
That is, every object that was live in \lean{HG} and is reachable from the roots
\lean{R}, from the constants of a live function, or is immortal is still live in
\lean{HG'}. The collector's specification \lean{gc_spec} takes the callee,
the VM stack and the globals as roots, and returns a new epoch
\lean{HG'} with \lean{epT HG HG' R}. Every continuation that may be resumed
after a collection, such as a \code{vm\_call} caller's or the JIT code's
\lean{jKont}, quantifies over the epoch it resumes in, and carries its own
roots (\lean{vstk B rs}) across the call. The interpreter switches epochs
at exactly one point, the safepoint at the beginning of \code{vm\_call}.
The collector's proof is split into marking (with an invariant on the mark
stack and the marked set), the root loops, draining, sweeping, which
frees the unmarked strings, arrays and objects, and the construction of the
new epoch.

\subsection{The bytecode compiler}
\label{sec:proof-compiler}

The safety of the interpreter and the JIT depends on the bytecode being well
formed: the interpreter does not check stack depths or operand ranges at run
time. The proof therefore splits into a predicate on bytecode and two
theorems: the compiler establishes it, and the interpreter and JIT are safe
for any code that satisfies it. \lean{CodeWF} requires
\begin{itemize}
\item every instruction start (a byte with a recorded depth $\ge 0$) to hold a
  valid opcode whose operands lie within the code and have no recorded depth;
\item operands to be in range: constant indices below \code{nconsts}, local
  indices below \code{nlocals}, global indices below 4096;
\item the recorded depth \code{depth[pc]} to be consistent along every edge:
  a jump's target and an instruction's fall-through successor have the depth
  that the instruction leaves, so control cannot fall off the end of the code;
\item no instruction to underflow the stack, and both the depth before and
  after an instruction to be at most \code{max\_stack} $\le 30{,}001$;
\item execution to start at $\mathit{pc} = 0$ with an empty stack.
\end{itemize}
A second predicate, \lean{WalkWF}, says that decoding the code linearly
from 0, as the JIT does, meets only instruction starts, including in
unreachable code.
The compiler writes \code{depth[pc]} and the jump-target flags as it emits, rather than running a
separate analysis, and the proof tracks a model of the emitted code
through the recursive-descent parser. The compiler's
specification \lean{cp_spec} says that \code{compile\_program}, given any byte string
as source code, either exits with a syntax error or returns a valid function
object, so \lean{CodeWF} holds for every function it creates. With
38{,}457 lines, the compiler proof is the largest single component other
than the JIT. One reason is that the lexer, the parser and the emitter are
all intertwined in one pass over the source.

\subsection{The interpreter}

The interpreter's specification \lean{VC} (\cref{fig:vc}) says that
\code{vm\_call(callee, args, argc)}, called with a valid callee and valid
arguments on the VM stack, returns a valid value or exits through a JavaScript
error, within a native stack budget that depends on the current call depth
\lean{d}.
The interpreter calls the JIT through three entry points: \code{jit\_compile},
\code{jit\_has\_entry} and \code{jit\_enter}. Their specifications form the
proposition \lean{JH}, and the interpreter is proved assuming it:
\[
\mlean{vm_call_spec} : \;\vdash \always \mlean{JH} \wand \mlean{VC}.
\]
The proof is by Löb induction, since \code{vm\_call} is recursive: the
\code{CALL} opcode handler calls \code{vm\_call}, and the induction hypothesis is
available under a later, which the call instruction strips. It covers every
opcode handler, the dispatch loop, argument setup, tier-up, OSR entry,
resumption after a deopt, and the calls to natives. JIT code can call back
into \code{vm\_call}, so \lean{JH}'s specification of \code{jit\_enter} in turn
takes $\later\mlean{VC}$ as a hypothesis (\cref{sec:jit}). Because \lean{JH}
assumes \lean{VC} only in this form, the JIT proof
\lean{jh_holds : textCode ⊢ JH} does not depend on the interpreter's proof,
and \lean{vc_holds : textCode ⊢ VC} combines the two.

\begin{figure}
\begin{lstlisting}[basicstyle=\ttfamily\scriptsize]
def VC : IP := □ ∀ (HG : HeapGS) x f nj d n Φ pm E ob vs,
  ⌜x 0 = 0 ∧ x 3 = gpAddr ∧ (x 2).toNat % 16 = 0 ∧
    8 * vcW nj d ≤ (x 2).toNat ∧ nj ≤ 1000 ∧ d ≤ 20000 ∧
    x 12 = vs.length ∧ ...⌝ -∗
  □ rtCtx pm -∗ □ Φ 1 -∗ □ vmRo E ob -∗ □ valOk (x 10) -∗
  vstk (x 11) vs -∗ slots (x 11 + 8 * vs.length) n -∗
  slots (x 2 - 8 * vcW nj d) (vcW nj d) -∗ vmWorld nj d -∗
  vcKR x f nj d (vs.length + n) Φ -∗
  WP (AExpr.run ⟨&vm_call#64, x, f⟩) {{ Φ }}
\end{lstlisting}
\caption{The specification of \code{vm\_call} (simplified). \code{a0} (\code{x 10}) is
the callee, \code{a1} the arguments \code{vs} on the VM stack, and \code{sp}
(\code{x 2}) must have room for \code{vcW nj d} words. \code{vmWorld} is the
runtime world (allocator, heap, kernel state, globals), and \code{vcKR} is the
continuation, which receives the result in a new epoch. Exiting through a
JavaScript error satisfies \code{$\Phi$ 1}.}
\label{fig:vc}
\end{figure}

\paragraph{Native stack.}
Stack overflow is a crash, so every specification accounts for native stack
usage: \lean{slots (x 2 - 8 * w) w} gives a function $w$ words of stack. The
interpreter's budget \lean{vcW nj d} depends on the JS call depth \lean{d}
(limited to 20{,}000) and on \lean{nj}, the number of arrays that
\code{to\_string} is currently joining (at most 1000, to detect cycles). The C code
checks these limits before recursing, and the binary's static 64\,MiB stack is
large enough for the worst case. The libc's start-up code switches to this
stack, so the kernel's stack limit does not matter.

\subsection{Putting it together}

The top-level module \code{Sep/Safe.lean} links all the specifications and
combines the start-up proof, \code{main}'s option parsing (including every
debugging flag), file reading, \lean{cp_spec} and \lean{vc_holds}. The result is
\lean{memory_safe : BinarySafe}. Finally, \code{ElfCheck.lean} proves by
\lean{decide +kernel} that the ELF loader, applied to the raw file, yields
exactly the entry point and segments of the generated image. This step takes
about 80~seconds, comparing the data word by word. It gives
\lean{memory_safe_elf}.

\section{Verifying the JIT}
\label{sec:jit}

The JIT is the reason for the project, and its proof is the largest part:
51{,}095 lines on RISC-V and 101{,}323 on x86. This section explains how it is
structured. The central difficulty is that the JIT's output is not a fixed
piece of code. It depends on the function's bytecode, which the attacker
controls, on the type feedback, which the attacker also controls, and on the
JIT's own analyses. We must show that \emph{every} piece of code that
\code{jit\_compile} can emit is safe, and that the binary's \code{jit\_compile}
emits only such code.

\subsection{Two layers}

We split the problem with a Lean model of the compiler.
\begin{enumerate}
\item \lean{Jit.compile : JIn → JS} is a functional transcription of
  \code{jit\_compile}, function by function, in a state monad over a
  record that mirrors C's \code{struct J}: the instruction buffer, the
  bytecode-to-code \code{pc\_map}, the jump fix-up list, the abstract types of
  locals and stack slots, the pending (virtual) stack, and a failure flag.
  The input \lean{JIn} holds the bytecode, constants, jump targets, the
  feedback bytes and the loop facts. C's \code{int}s become \lean{Int}s, loops
  become structural recursions, and instruction words are built by the same
  bit formulas as the C encoders.
\item \textbf{Code safety}: for every well-formed function and \emph{every}
  input, if \lean{compile} does not fail then the code it produces, placed at
  any address, is safe to enter at offset 0 and at every jump target that
  the \code{pc\_map} lists.
\item \textbf{Refinement}: the binary's \code{jit\_compile}, run on a function
  object, writes into fresh executable memory exactly the words of
  \lean{(compile i).buf} and the \code{pc\_map} of \lean{compile i}, where
  \lean{i} is the function's current bytecode and feedback, together with
  the loop facts that a first pass computes (\lean{jit_compile_spec}).
\end{enumerate}
The model is not trusted: the refinement proof connects it to the binary.
Still, a model that disagrees with the C code makes the refinement proof fail
late and in a confusing way. We therefore test it differentially:
\code{tests/jitmodel/run.sh} dumps every compilation of the test suite and the
benchmarks from an instrumented build, replays each one through
\lean{compile} compiled by Lean, and checks that the code words, the pc maps and
the failure flags are identical. There is one such test for each architecture.

The interface between the JIT and the rest of the proof is the class
\lean{JitModel}, which the heap invariant uses to describe \code{fn->jit}:
\begin{lstlisting}
def jitCode (p : Addr) (fi : FnInfo) (j : Addr) : IP :=
  ∃ (i : J.In) (mem pmA : Addr), ⌜J.ok fi i ∧ ...⌝ ∗
    ptsN j .discard rwPerm 8 mem ∗ ptsN (j + 16) .discard rwPerm 8 pmA ∗
    J.code mem i ∗ □ (∀ pc, ⌜pc < fi.fc.len⌝ →
      ptsN (pmA + 4 * pc) .discard rwPerm 4 (J.pcMap i pc))
\end{lstlisting}
\noindent
A function's JIT code is a persistent fact: the code \lean{J.code mem i} (for
RISC-V, \lean{codeAt mem (compile i).buf}, that is, \lean{instrAt} for each
word) and its pc map are read-only forever. JIT code is never freed: when a
function deoptimizes, \code{fn->jit} is cleared, but other activations may still
be running the old code. The heap ghost state, the interpreter and the GC
work with any \lean{JitModel}, which is how they are shared between the two
architectures.

\subsection{Installing code}

\code{jit\_compile} compiles into a \code{malloc}ed buffer, then maps fresh
pages with \code{mmap} (read-write), copies the code, and calls
\code{mprotect} to make them read-execute. On RISC-V it then calls
\code{riscv\_flush\_icache} and, if any step fails, unmaps the pages and
returns \code{NULL}. In the logic, \code{mmap} returns \lean{bytesAt} with
write permission, \code{mprotect} converts it to a discarded (read-only)
fraction with execute permission, and only the flush rule produces the
\lean{cleanAt} tokens that \lean{instrAt} needs. This is W$\oplus$X: code pages are
never writable and executable at the same time, are never made writable
again, and are never patched. That is why read-only ownership of code can be
persistent and duplicable: the code can be shared by any number of
activations, nothing can write to it, and the proof needs no reasoning about
self-modifying code.

\subsection{The safety invariant}

Code safety is stated for a fixed environment \lean{ε : JEnv}: the model
input, the code address, the frame, the VM stack area, the registers at entry
and the continuation. The persistent part is \lean{jCtx ε Φ}: the model did not
fail, the code is at \lean{ε.B}, the function object is live, and the various
layout facts hold. The ownership part, \lean{jRes ε Φ ls st}, owns the
interpreter frame holding the locals \lean{ls} and the operand stack \lean{st},
the unused slots of the frame, the runtime world, the saved registers and the
native stack, and the continuation \lean{jKont}, which is \code{jit\_enter}'s
postcondition. All values in \lean{ls} and \lean{st} satisfy \lean{valOk}.

\paragraph{Register-allocated locals.}
On RISC-V, the first five locals live in callee-saved registers. Their
frame slots are stale. \lean{jRes} therefore owns a frame holding
\lean{lm}, which agrees with \lean{ls} except at the register locals, and the
pure fact \lean{RegsOk ε x ls} says that the registers hold the current values.
Every call out of JIT code preserves them because they are callee-saved, which the
callee's spec states through \lean{Clobbers}. Before a call that can trigger a
GC, the code spills the register locals to the frame, so they are roots.

\paragraph{The virtual stack.}
The RISC-V JIT delays pushes: a constant or a local that is pushed and
consumed by the next instruction is never stored. The compiler state records
for each slot above the \emph{pending base} \lean{pb} whether it is in a
register, a local, or a constant. The invariant \lean{jResP} says that the slots
from \lean{pb} upward are where the compiler state says they are, and that
memory holds only the stack below \lean{pb}.

\paragraph{Type facts.}
For guard elimination, the compiler tracks an abstract type for each local and
stack slot, and whether a slot is a copy of a local. The facts are an
Iris proposition:
\begin{lstlisting}
def tyV (t : Nat) (v : BitVec 64) : IP :=
  (if t = AT_ARRAY then arrV v else emp) ∗ ⌜t = AT_BOOL → v = V_TRUE ∨ v = V_FALSE⌝
def tyOk (s : JS) (ls st : List (BitVec 64)) : IP :=
  ⌜∀ k < st.length, 0 ≤ s.ssrc[k] → st[k]? = ls[s.ssrc[k]]?⌝ ∗
  □ (∀ k v, ⌜st[k]? = some v⌝ → tyV s.stype[k] v) ∗
  □ (∀ j v, ⌜ls[j]? = some v⌝ → tyV s.ltype[j] v)
\end{lstlisting}
\noindent
Note what \lean{tyV} leaves out. Integer and double facts say nothing,
because they are irrelevant to safety: code that wrongly treats a value as an
integer computes with garbage bits, but it re-boxes the result into a valid
integer. An array fact, on the other hand, means that the JIT reads the
array's length and data pointer without a check, so it must be backed by
\lean{objAt}, and boolean facts matter because the code computes \code{!b} by
flipping its low bit. The safety invariant is weaker than a full typing invariant,
which made the proof smaller. The facts must survive garbage collection, since
array facts mention \lean{objAt} in the current epoch. \lean{tyOk_epoch}
transports them along \lean{epT}, provided that every value in the frame is a root.

\paragraph{The main loop.}
The compiler walks the bytecode linearly: the trajectory \lean{traj i m} is the
bytecode pc and compiler state after $m$ steps. The specification
\lean{SS ε Φ m} says that the code emitted at step $m$ is safe, starting
from any register file and frame consistent with the compiler state before
step $m$:
\[
\begin{array}{l}
\always \forall x\,f\,\mathit{ls}\,\mathit{st}.\;
  \pure{\mathit{pc}_m < \mathit{len} \land \mathit{depth}(\mathit{pc}_m) = |\mathit{st}| \land {}\\
  \qquad \mlean{RegsOk}~\varepsilon~x~\mathit{ls} \land \ldots} \wand
  \mlean{tyOk}~s_m~\mathit{ls}~\mathit{st} \wand {} \\
\quad \mlean{jResP}~\varepsilon~\Phi~s_m~x~\mathit{ls}~\mathit{st} \wand
  \later\mlean{VC} \wand {} \\
\quad \wpre{\langle B + 4|s_m.\mathit{buf}|, x, f\rangle}{\Phi}
\end{array}
\]
\lean{SS_all : jCtx ε Φ ⊢ ∀ m, SS ε Φ m} is proved by Löb induction together with a
downward induction on $m$, from the end of the trajectory. A step's code
continues at the next step, which the downward induction provides, or jumps to an arbitrary step,
which the Löb hypothesis provides under the later of the jump instruction.
The Löb induction ranges over all heap epochs, because
the code of a call can resume in a new epoch.

The proof of one step is a case split on the opcode and the code-generation
strategy, which is chosen from the feedback byte. Each case is a
\emph{template lemma}: the instructions emitted for, say, an integer addition
with an overflow check, starting from a compiler state $s$, run from a frame
satisfying \lean{tyOk s} and \lean{jResP} to the end of the template's code
with a frame satisfying the invariants for the state after the step, or they
branch to the instruction's deopt stub. The template lemmas are stated over the
\emph{output} of the model's emitters, such as \lean{exec (genInsn pc op d) i s},
so the arithmetic on offsets, register numbers and immediates is evaluated
symbolically, and the lemmas hold for all of them at once. The
per-opcode proofs are in about 100 modules, which build in parallel.

\paragraph{Jumps and OSR.}
At a jump target, the compiler forgets all type facts and materializes the
virtual stack, so the invariant at a target is the trivial one
(\lean{tyOk_pre_tgt}). Any jump can thus continue there. OSR entries are
exactly these points: \code{jit\_enter(jc, f, pc)} with a pc whose
\code{pc\_map} entry is set runs the prologue, which saves the
callee-saved registers and loads the register locals, and then jumps to the
instruction's code. Its precondition \lean{frameAt} is the interpreter's
frame invariant, which implies \lean{jRes} with no type facts.

\paragraph{Loop facts.}
Loop facts complicate this picture. A loop header with facts starts with
checks of the assumed local types, and the code after the checks, the head,
knows them. A backward jump whose state implies the facts goes straight to
the head and skips the checks. All other entries, including OSR, go through the
checks. The facts \lean{lf} are an input of the model, and the proof
holds for every \lean{lf}: the C code computes them in a first pass, but
safety does not depend on how. Hence the loop-fact analysis is not verified
and does not need to be. If it is wrong, the code deoptimizes.

\paragraph{Exits.}
Each instruction's guards branch to a shared deopt stub after the
instruction, which materializes the virtual stack as it was at the start of the
instruction, loads the bytecode pc into \code{t0} and the stack depth into
\code{t1}, and jumps to the common exit. The exit spills the register locals,
stores pc and depth into the frame, restores the callee-saved registers, and
returns 1 (\lean{wp_stub}, \lean{wp_exit}). The interpreter resumes there,
re-executing the instruction that failed. For this to be safe, the frame must
satisfy the interpreter's invariant \lean{frameAt} at that pc. This holds
because the JIT's frame always contains valid values and has the depth that
\lean{CodeWF} assigns to the pc. Since the instruction has not had any
visible effect when a guard fails, the frame is the one before the instruction.
Note that this proof says nothing about whether the deopt is \emph{correct},
that is, whether the interpreter computes the same result that the JIT would
have; only that the interpreter can safely continue.

\paragraph{Calls.}
JavaScript calls from JIT code go to the runtime's \code{rt\_call}, which calls
\code{vm\_call}; JIT code is thus re-entrant. The specification of
\code{jit\_enter} assumes $\later\mlean{VC}$, and the JIT proof uses it for the call
helper. Calls to other helpers (\code{rt\_binop}, \code{rt\_get\_prop}, and so on)
use their specifications directly. Before each call, the code writes the
register locals and the virtual stack back to the frame, which is also what
makes them roots for the GC. After the call, the template lemma moves the type
facts and the continuation to the new epoch.

\subsection{Refinement: \code{jit\_compile} computes \lean{compile}}

The refinement proof is a conventional, if large, separation-logic proof of a
C program with \code{malloc}, \code{realloc}, \code{free} and loops: each C function
in \code{jit.c} has a specification saying that, on a buffer holding
the model state $s$, it leaves the buffer holding the state after the
corresponding model function. Because the model is a literal transcription,
each function's proof follows its code. The arithmetic of the model is on
unbounded integers, so the proof also shows that the C arithmetic does not
overflow, using the bounds of \lean{CodeWF}. \code{jit\_compile} runs up to three passes (a first pass to
compute the loop facts, the real one, and a fallback without facts if that
fails), and may return \code{NULL}, since \code{mmap}, \code{mprotect} or the
flush can fail. Running out of heap memory is a clean exit. Finally,
\lean{jh_holds} packages \lean{jit_compile_spec}, the specification of
\code{jit\_has\_entry} and that of \code{jit\_enter} (which runs the prologue and
then uses \lean{SS_all}) into \lean{JH}.

\subsection{What made this tractable}

Three choices kept the JIT proof manageable. First, the two layers separate
two very different kinds of reasoning, about the semantics of generated code
and about a C program manipulating buffers, and each can be developed and
changed independently, since they meet only at \lean{compile}. Second, the JIT
was designed for the proof: one compiler state, deopt stubs per instruction
rather than per guard, and type facts reset at every jump target. Third, the
invariant is about safety only. The proof never has to say which value the
JIT code computes, only that it is valid, and this is what makes it possible
to prove the code safe for any feedback.

\section{The x86-64 Port}
\label{sec:x86}

To test whether the approach depends on the simplicity of RISC-V, we ported
the engine and the proof to x86-64. The x86 backend (\code{src/x86\_64/jit.c},
1{,}063 lines) is a port of the RISC-V JIT that keeps its design: the same
frame layout, strategies, abstract types, guard elimination and deopt
stubs, but no register allocation, virtual stack or loop facts. The C code
is otherwise shared, and the binary is built with gcc's default
\code{-march=x86-64}. The x86 proof states the same theorem,
\lean{memory_safe_final : BinarySafe}, over the Sail x86-64 model, for every
command line including the debugging options. One link is weaker than on
RISC-V: there is no x86 counterpart of \lean{memory_safe_elf}. The theorem is
about \lean{Image.entry} and \lean{Image.segments}, which the script
\code{proofs/tools/elf2lean\_x86.py} extracts from the executable, so on x86-64 that
script is part of the trusted base. It took nine days from the
first commit on the x86 model to the unconditional theorem, plus two days
for the debugging options.

\subsection{The model}

We generate Lean from the sail-x86 model~\cite{sail-x86}, which had not been
compiled to Lean before. This required patches to the model and a
post-processing script, because we hit ten bugs in Sail's Lean backend
(for example, \code{return} inside \code{try}/\code{catch} handlers,
unescaped Lean keywords used as identifiers, and record updates that do not
parse) and two problems with Lean~4.32: its new \code{do} elaborator fails
on some join points followed by loops, and some recursive Sail functions
have no termination measure. The model also needs assumptions that the RISC-V one
does not:
\begin{itemize}
\item \textbf{CPU profile.} The processor is pinned to x86-64-v1 (SSE2, no
  later extensions), by turning the CPUID feature flags into constants. This
  is what gcc and our JIT emit, and the engine never executes \code{cpuid}.
\item \textbf{Floating point.} The model's 162 floating-point externs are
  opaque constants. This is sound as long as MXCSR keeps its default value
  (round to nearest, exceptions masked), and the binary contains no
  instruction that changes it.
\item \textbf{Loops.} Sail \code{while} loops are compiled to Lean's
  \code{Loop} type, which is unspecified in the free monad. We give a
  fueled instance that runs at most $2^{64}$ iterations and then fails an
  assertion. A \code{rep movsb} with a 64-bit count terminates within this
  bound.
\end{itemize}
The process model is the RISC-V one, minus \code{riscv\_\allowbreak flush\_\allowbreak icache}. On
x86 Linux, a single thread sees its own stores in instruction fetches, so
there is no dirty-byte tracking. The address space follows Linux's default
for 4-level paging: \code{mmap} results and the initial stack lie below
$2^{47}-4096$ (\cref{sec:findings}). The \code{mprotect} and \code{munmap}
contracts both require the range to be mapped, whereas RISC-V, like Linux,
also allows \code{munmap} of unmapped pages; this is a stronger obligation on
the program. A processor exception is recorded by
the model in a pending-fault flag rather than delivered. Writing this flag,
changing the privilege level or mode, or failing an assertion has no step.
A \code{syscall} instruction is executed atomically by the kernel.

\subsection{Bridging variable-length instructions}

On RISC-V we decode \code{.text} into a list of instructions. x86
instructions have variable length, so the image is instead a list of pages
of raw bytes. The proofs cut a function's bytes out of \code{.text} with
\lean{decide +kernel}, and the model decodes them. A bridge lemma covers one
encoding, with the ModRM byte, the REX prefix and the operands symbolic, and
gives the new values of RIP, the general-purpose registers and the flags, and
the exact list of memory events. The flags are stated as bit-vector
operations, for example \lean{BitVec.saddOverflow} for OF.
There are 474 bridge lemmas in 146 files. Most of them are generated by
scripts (ALU forms by operation, operand size and prefix; loads and stores by
addressing mode; SSE moves, arithmetic, comparisons and conversions; shifts;
condition codes), and the rest are written by hand. A coverage script
disassembles the binary and reports every instruction that no lemma matches.
Some instructions needed special care:
\begin{itemize}
\item \code{idiv} traps on a zero divisor and on \code{INT\_MIN / -1}; its
  lemma requires the divisor to be nonzero and the quotient to fit, and the
  JIT checks the divisor and divides in 64 bits so that the second case
  cannot occur.
\item \code{rep movs} and \code{rep stos} are loops in the model; their
  lemmas are proved by induction over the fueled loop with an invariant per
  iteration.
\item gcc inlines \code{fmod} using the x87 \code{fprem} instruction, so
  there are lemmas for a few x87 instructions, with hypotheses on the x87 tag word that
  rule out stack faults.
\item \code{ucomisd} sets ZF, PF and CF for unordered operands, so the JIT
  chooses comparisons for which NaN gives false. x86 arithmetic
  propagates NaN payloads, so a result can collide with a NaN-box tag, and
  the JIT checks every floating-point result
  (\cref{sec:engine}). The default NaN \code{0xFFF8\ldots} is itself a valid
  double and is left alone.
\end{itemize}

\subsection{Sharing the proof}

The x86 program logic is the same Iris instantiation, with the
\lean{codeAt a bs} resource for byte ranges instead of instruction words. The
architecture-neutral modules (the ghost state, points-to, byte strings, stack
slots) are shared by both developments. The models of the runtime data
structures (the allocator's size classes and \lean{mblock}, the heap and
value invariants, the shapes and inline caches) are also shared, and they
refer to global addresses only through a class \lean{Layout} and to JIT code
only through \lean{JitModel}, which each target instantiates.

The function proofs themselves are not shared, since they are about different
machine code. The structure is the same as on RISC-V, however, and so are the
specifications, up to the calling convention, so the RISC-V proofs served as a
blueprint. The x86 JIT has its own Lean model (\lean{x86Jit}), its own differential
test, and the same two-layer proof. The x86 JIT proof is larger than the RISC-V
one (101k lines vs.\ 51k), in part because an x86 instruction does more than a
RISC-V instruction: a single \code{add} with a memory operand reads memory,
computes and sets six flags, and the proofs must track the flags that later
branches depend on.

\section{What the Proof Found}
\label{sec:findings}

\safejit was written to be verified, and it was tested and hardened before
the proof started: a differential test suite against Node.js, a robustness
suite of about 200 malformed and adversarial programs, and a round of review and
fuzzing, which led to fixes for deep nesting, cyclic arrays, out-of-bounds reads
in the lexer, length overflows, and unchecked \code{realloc} and
\code{mprotect} results. Nevertheless, two bugs that testing had not found surfaced during
verification. Neither was found by the proof alone. Both
were found when we asked an agent based on a different model to
assess how strong the safety claim was and where its assumptions were
unrealistic. That review pointed out two places where the trusted model was too
optimistic, the instruction cache and the size of the address space. Once we made the
model faithful, the proofs of the existing code failed, and the failures
pointed at the code that had to change.

\paragraph{Unchecked instruction-cache flush.}
The Sail RISC-V model executes \code{fence.i} as a no-op, and the first
version of the process model did not model instruction caches at all. With that
model the JIT was provably safe, but the model was wrong: a real RISC-V core
can fetch stale instructions after the JIT writes new code. We added dirty-byte
tracking to the process model (\cref{sec:theorem}), and the proof of
\code{jit\_compile} failed. (The review had already predicted this failure.) The C code called gcc's
\code{\_\_builtin\_\_\_clear\_cache}, which on Linux/RISC-V calls the
\code{riscv\_flush\_icache} system call, but discards its result. If the
system call fails, the engine runs code that the hart may not see. The fix is to
call \code{\_\_riscv\_flush\_icache} directly and unmap the code and fail
the compilation if it returns an error. The system call rarely fails in practice,
but this is exactly the kind of assumption that verification is supposed to
eliminate: the contract says that it may fail, so a proof must handle it.

\paragraph{Addresses above $2^{47}$.}
\safejit's NaN-boxing stores pointers in the low 48 bits of a value. The first
version of the trusted model assumed that the kernel places all memory below
$2^{38}$, which is the user address limit on Sv39. That is not true on
Sv48, where it is $2^{47}$, nor on Sv57: there, before Linux~6.6, the initial
stack could be placed just below $2^{56}$, and \code{mmap} returns addresses above
$2^{47}$ once the lower part of the address space is exhausted. A pointer above
$2^{48}$ would be truncated when boxed, and the engine would access the wrong
memory. When we changed the model to let the kernel use any address below $2^{56}$,
the proofs of the start-up code and of \code{mmap}'s callers failed. Now
\code{\_\_libc\_start} exits with status 127 if the initial stack or any
\code{argv} string lies at or above $2^{47}$, and the \code{mmap} wrapper
unmaps any mapping that reaches above $2^{47}$ and fails with
\code{ENOMEM}, so that \code{malloc} and \code{jit\_compile} fail cleanly.
The checks are in the shared libc, so the x86-64 build has them too. The
x86-64 model, however, follows Linux's default 47-bit address space with
4-level paging: its \code{mmap} contract and initial stack stay below
$2^{47}-4096$, so its proof steps through the checks but never has to handle
a successful high-address allocation or a high initial stack.

\paragraph{A lesson about the trusted base.}
Both bugs survived a mechanized proof for a while because the trusted model
encoded the same optimistic assumption that the programmer (here, an agent)
had made. The trusted base deserves as much review as the code. For each
feature of the real system that the model abstracts away (caches,
address-space layout, failing system calls), one should ask whether the
abstraction is conservative. Asking an agent built on a different
model was a cheap way to get such a review, and it found
both problems in minutes. We now state the
syscall contract so that every call may fail, every \code{mmap} may return any
address below $2^{56}$, and only a successful flush makes new code fetchable.

\paragraph{Tools.}
Generating the models exposed bugs in the tools as well. Sail's Lean backend printed
\code{flen = flen\_bytes * 8} without parentheses in the RISC-V model, making
every floating-point register eight bits wide. The x86 model hit
ten more bugs in the Lean backend and two problems in Lean
(\cref{sec:x86}). Such bugs are a reminder that the generated model is part of the trusted base:
a model bug that makes an instruction behave more nicely than the hardware could
hide a real bug in the engine.

\paragraph{Engine changes for verifiability.}
Some changes were not bug fixes but made the engine easier to verify. The
compiler records stack depths and jump targets as it emits code, replacing a
separate bytecode verifier that ran at load time; the proof that the
compiler's output is well formed is simpler than a proof of the verifier would
have been, and it removes a pass. Every code generator in the JIT is marked
\code{noipa}, so that gcc compiles it separately, and a change to one generator
does not change the machine code, and hence the proofs, of the others. The
engine is compiled for RV64G without compressed instructions, so that
instructions are aligned 32-bit words. The libc is freestanding, with raw
system calls and a static 64\,MiB stack, so that the specifications
need not depend on glibc or on resource limits.

\section{Evaluation}
\label{sec:eval}

We evaluate three things: whether the JIT is worth having (performance),
how the engine was tested beyond the proof, and what the proof costs (size and
checking time).

\subsection{Performance}

\safejit is a baseline JIT, and its benchmarks are small kernels that exercise
the four things it optimizes: calls (\code{fib}, a naive recursive Fibonacci
of 27), integer loops (\code{loop}, five million iterations of an
integer update), floating-point loops (\code{mandel}, a $160\times160$
Mandelbrot set with at most 200 iterations per point) and array accesses
(\code{sieve}, ten runs of the sieve of Eratosthenes up to 200{,}000).
\Cref{tab:bench} shows the running times. The RISC-V build runs under
user-mode \code{qemu-riscv64} on an x86-64 host, so its absolute times mean
little, but the speedup of the JIT over the interpreter is meaningful: both
run under the same emulator. The x86-64 build runs natively on an AMD EPYC~7V12.
We report the best of ten runs, measured end to end; \safejit's startup is
about half a millisecond.

\begin{table}
\caption{Running times in seconds (best of 10). RISC-V under qemu-user,
x86-64 native. Node.js~18 times include about 0.09\,s of startup.}
\label{tab:bench}
\small
\setlength{\tabcolsep}{4pt}
\begin{tabular}{@{}l rrr rrr rr@{}}
\toprule
& \multicolumn{3}{c}{RISC-V (qemu)} & \multicolumn{3}{c}{x86-64} & \multicolumn{2}{c}{Node.js} \\
\cmidrule(lr){2-4}\cmidrule(lr){5-7}\cmidrule(l){8-9}
& interp & JIT & $\times$ & interp & JIT & $\times$ & jitless & full \\
\midrule
fib    & 0.29 & 0.13 & 2.2  & 0.037 & 0.014 & 2.6 & 0.141 & 0.117 \\
loop   & 7.75 & 2.27 & 3.4  & 0.438 & 0.148 & 3.0 & 0.301 & 0.141 \\
mandel & 5.10 & 0.22 & 23.1 & 0.362 & 0.039 & 9.3 & 0.250 & 0.119 \\
sieve  & 3.77 & 0.15 & 25.3 & 0.459 & 0.057 & 8.1 & 0.373 & 0.149 \\
\bottomrule
\end{tabular}
\end{table}

On RISC-V, the JIT is 2--25$\times$ faster than the interpreter. The gains are
largest where speculation removes the most work: in \code{mandel} and
\code{sieve}, loop facts let the JIT drop the type checks on locals, and
doubles and array elements are handled inline instead of through runtime
helpers. \code{fib} gains least, because every call goes through the
runtime's \code{vm\_call}, which sets up an interpreter frame. The x86-64 JIT
has no register allocation, virtual stack or loop facts (\cref{sec:x86}), and
every local lives in memory, so its speedups are smaller, 2.6--9.3$\times$.
(Speedups under an emulator need not match those on RISC-V hardware, which we
did not measure.)

To put the numbers in context, \cref{tab:bench} also shows Node.js~18 on the
same programs (with a small prelude that defines \code{print}). End to end, \safejit's
x86-64 JIT is faster than Node on \code{fib}, \code{mandel} and \code{sieve}
and about as fast on \code{loop}, and V8's interpreter
(\code{--jitless}) is slower than \safejit's JIT on all four. But Node's times
include about 0.09\,s of startup, which dominates runs this short, while
\safejit starts in under a millisecond. Subtracting the startup gives only a
rough estimate of V8's optimizing tiers: about 2.9$\times$ faster than
\safejit's x86-64 JIT on \code{loop}, 1.3$\times$ on \code{mandel}, on par on
\code{sieve}, and about 2$\times$ slower on \code{fib}, where the startup is
most of Node's time. A precise comparison would time the kernels in-process.
We do not claim more than that \safejit's JIT is a real JIT: it gives the kind of
speedup that justifies a baseline tier, on the kinds of code that baseline tiers
are for.

\subsection{Testing}

The proof says nothing about functional correctness (\cref{sec:theorem}), so
the engine is tested as well:
\begin{itemize}
\item \textbf{Differential tests against Node.js.} Eleven test programs
  (arithmetic, arrays, bit operations, calls, comparisons, cyclic structures,
  doubles, errors, loops, objects, strings) are run in four modes: interpreter only, the default threshold of 200, compile on the
  first call (\code{--jit-threshold=1}), and compile on the second call while
  allowing up to 1{,}000 deopts (\code{--jit-threshold=2 --max-deopts=1000}),
  which compiles with type feedback from a single call and so deoptimizes
  frequently. Each output is compared with Node's. The benchmark script also
  checks that the interpreter and the JIT print the same results.
\item \textbf{Robustness.} A generator produces 204 hostile inputs: nesting
  200{,}000 levels deep in every syntactic position, operand stacks near and
  beyond their limits, lexer edge cases, huge strings and arrays, and 150 random
  token sequences. Each runs in all four modes, and the test fails if any run
  is killed by a signal or times out. This suite, and the fixes it drove, predate
  the proof.
\item \textbf{libc.} The freestanding \code{printf}, \code{strtod} and math
  functions are tested against glibc.
\item \textbf{The JIT model.} An instrumented build of the RISC-V engine logs
  every compilation (bytecode, feedback, emitted code). A Lean program runs the
  model \lean{Jit.compile} on each logged input and checks that it produces the
  same code byte for byte. This test was the main tool for writing the
  model (\cref{sec:jit}). The x86-64 model is tested the same way.
\end{itemize}

The proof makes part of the robustness suite redundant: no input can
crash the engine, since a crash is a stuck state (a timeout is still possible). But the tests are cheap, and
they were invaluable during development, when a failing differential test
pointed at a bug in minutes that would have taken hours to find from a stuck
proof.

\subsection{Proof size and cost}

\Cref{tab:size} shows the size of the development. The RISC-V proof is about
221k lines of Lean, of which 57k are generated: the dump of the ELF image and
its symbols, and the instruction datatype. The generated Sail models are not
counted. The x86-64 proof is about 264k lines, of which 12k are generated: the
image, the symbols and the bridge lemmas produced by scripts. In total, the
two proofs are about 485k lines of Lean for 6.5k lines of C, a ratio of
about 75:1 (64:1 excluding generated code). For the initial version (4.4k lines
of C and 194k lines of Lean), the ratio was 45:1; the JIT optimizations added
after it (register allocation, the virtual stack, loop facts) are where the
proof grew fastest.

\begin{table}
\caption{Size in lines. Indented rows are included in the row above;
\emph{generated} is included in the total. \emph{Sep} is the
separation-logic layer with the program proofs; \emph{Jit} is the JIT model
and its proofs. The generated Sail models are not counted.}
\label{tab:size}
\small
\begin{tabular}{@{}lr@{\hspace{2em}}lr@{}}
\toprule
\multicolumn{2}{@{}l}{C (engine + libc)} & \multicolumn{2}{l@{}}{Lean, RISC-V} \\
\midrule
engine (\code{src})   & 3{,}868 & \emph{Sep}  & 103{,}540 \\
\quad RISC-V \code{jit.c} & 1{,}376 & \quad compiler & 38{,}457 \\
libc                  & 1{,}545 & \quad interpreter & 7{,}870 \\
x86-64 \code{jit.c}   & 1{,}063 & \quad GC & 4{,}145 \\
                      &         & \emph{Jit}  & 51{,}095 \\
                      &         & process model & 7{,}748 \\
                      &         & ELF image, symbols & 55{,}783 \\
                      &         & other & 2{,}655 \\
\cmidrule{3-4}
                      &         & total & 220{,}821 \\
                      &         & \quad generated & 57{,}488 \\
\midrule
total C               & 6{,}476 & \multicolumn{2}{l@{}}{Lean, x86-64} \\
\cmidrule{3-4}
                      &         & \emph{Sep} & 138{,}829 \\
                      &         & \emph{Jit} & 101{,}323 \\
                      &         & bridges & 16{,}197 \\
                      &         & other & 7{,}466 \\
\cmidrule{3-4}
                      &         & total & 263{,}815 \\
                      &         & \quad generated & 12{,}022 \\
\bottomrule
\end{tabular}
\end{table}

A few steps dominate the time it takes to check the proofs. Compiling the Sail RISC-V model takes about 45 CPU-minutes. Each of
the floating-point bridge lemmas takes about five minutes, and they build in
parallel. Running a function's straight-line code symbolically
(\lean{twalk}) takes 5--25 seconds per cycle in the control-flow graph. Checking
that the ELF file matches the image (\lean{ElfCheck}, by \code{decide +kernel})
takes about 80 seconds. Since Lean builds are incremental at the module level, a change to one
function's code mostly re-checks that function's proof, which is why every code generator is compiled as a separate function
(\cref{sec:findings}).

\section{Development with AI Agents}
\label{sec:agents}

All of \safejit (the engine, the libc, the tests, the Lean models, the proofs and
the proof tooling) was written by AI agents: Claude Opus~5.5 running in the
GitHub Copilot CLI, guided by one researcher. Of the 1{,}411 non-merge
commits on the main branch before this report, 1{,}395 carry an agent's
session identifier; the researcher's own
commits are a handful of README edits and housekeeping. This section describes
how the work was organized, what it cost, and what we learned. We think
this is the most important result of the project: not that a small JIT can be
verified, which was never in doubt, but that it took four weeks.

\subsection{Setup and timeline}

The project started from a single prompt, asking for ``a simple javascript
engine with a risc-v jit compiler'' that does ``some basic type speculation''.
The agent produced the engine in under two hours. A second agent session was
asked to replicate the approach of xv6iris on this engine, with the
goal of proving it memory safe ``in particular that the optimized jit-generated
code doesn't break that safety either'', and allowing it to change how the
engine is built. This session ran for four weeks and produced the
RISC-V proof. \Cref{tab:timeline} shows the milestones.

\begin{table}
\caption{Milestones (2026). Commits per day peaked at 339, 344 and 250 on
Nov.~16--18, when parallel workers were active.}
\label{tab:timeline}
\small
\begin{tabular}{@{}lp{0.74\columnwidth}@{}}
\toprule
Oct 23 & Engine written; verification session starts \\
Oct 28 & Researcher review: proofs are not using separation logic; migration to Iris \\
Nov 6  & First unconditional safety theorem for the RISC-V binary \\
Nov 9  & Allocator with a verified \code{free}; initial version complete (404 commits) \\
Nov 10 & Objects and inline caches, GC; x86-64 session starts \\
Nov 13 & I-cache and address-space model fixes (\cref{sec:findings}) \\
Nov 14 & Debugging options verified \\
Nov 15 & RISC-V register allocation \\
Nov 16 & ELF cross-check; x86-64 proofs parallelized \\
Nov 17 & Loop facts \\
Nov 18 & Unconditional theorem for the x86-64 binary \\
Nov 20 & x86-64 debugging options verified \\
\bottomrule
\end{tabular}
\end{table}

The initial RISC-V version (engine, interpreter, baseline JIT with
speculation, OSR and deopts, allocator, and the proof) took about ten working
days. Later features (objects, the GC, register allocation, the virtual
stack and loop facts) were added in the same session, each on its own
branch, and merged once the proof covered it again.

For the x86-64 port, the researcher started a second, concurrent session.
For its first week, that session worked alone: it ported the engine,
patched the Sail x86 model's Lean output, generated the bridge lemmas and
built the shared modules. On Nov.~16 the researcher suggested porting the specification
files first and then discharging the proofs in parallel. The agent then acted as
a coordinator: it wrote the specifications, assigned groups of functions
to worker agents (22 sub-agents over the session, typically about six at a
time), each in its own branch, reviewed and merged their branches, and
forwarded fixes between workers. Within three days the x86-64 theorem
was proved without premises.

\subsection{What the researcher did}

\Cref{tab:effort} summarizes the interaction. Across all sessions, the
researcher wrote about 180 prompts totalling about 17{,}000
characters, roughly the size of this section. Of these, 68 were some
variant of ``Excellent work, please continue!''\ and another 14 asked the agent
to merge a branch. The rest fall into a few categories:
\begin{itemize}
\item \textbf{Setting goals.} The initial prompts, and the choice of each next
  feature: ``Are we already in a good shape to add a GC?'', ``let's tackle
  inferring type facts across loop back-edges next'', ``There's a brand new x86
  model for sail \ldots\ we should try it.''
\item \textbf{Steering the design.} The most consequential intervention came on
  the fifth day: ``We don't seem to be using separation logic at all. It's a whole
  lot slower than xv6.'' Until then the agent had reasoned about
  executions directly, with a whole-machine safety notion and no framing, and
  about 28k lines of proofs had to be migrated to the Iris-based logic. Others asked to prove properties instead of checking them at run time
  (which led to bytecode metadata by construction, \cref{sec:findings}), to make proofs
  robust against recompilation (which led to symbolic addresses), to keep the
  x86-64 proofs in separation logic rather than repeating a ``historic
  artifact'' of the RISC-V proof, and to align the x86-64 specification pattern with the RISC-V one.
\item \textbf{Keeping the agent honest.} The researcher asked the agent to remove a
  \lean{sorry} that was only a fallback in a tactic, and challenged a
  status report that described the RISC-V JIT proof as an assumption, which it was not. The researcher also
  asked a different agent how strong the safety claim was, which led to
  the two model fixes of \cref{sec:findings}.
\item \textbf{Operations.} Machine reboots, a stuck build, and file-descriptor
  limits that the parallel builds exceeded.
\end{itemize}
The researcher did not write proofs, and did not need to read most of
them: the Lean kernel checks them, and the only things a human must
check are the statement of the theorem, the trusted base, and the axioms
the proof depends on (\cref{sec:theorem}).

\begin{table}
\caption{Effort. Model time is the sum of the durations of model requests,
including concurrent ones. Tokens include cached input. The \emph{all} column
also includes the sessions for the initial engine, the adversarial review and
tooling, so it exceeds the sum of the other two.}
\label{tab:effort}
\small
\begin{tabular}{@{}lrrr@{}}
\toprule
& RISC-V & x86-64 & all \\
\midrule
calendar days & 29 & 11 & 29 \\
researcher prompts & 56 & 69 & 182 \\
sub-agents & 2 & 22 & 33 \\
model requests & 33.6k & 41.5k & 79.5k \\
input tokens & 3.6\,G & 4.4\,G & 8.4\,G \\
\quad of which cached & & & 97\% \\
output tokens & 33\,M & 34\,M & 71\,M \\
model time & 91\,h & 94\,h & 196\,h \\
\bottomrule
\end{tabular}
\end{table}

\subsection{Lessons}

\paragraph{The proof checker is the reviewer.}
A proof assistant is an ideal environment for an agent: every claim is either
checked or not, and the agent cannot talk its way past the kernel. Agents do
make mistakes in their reports: once, an agent missed a \lean{sorry} in a
lemma it had reported as proved (and found and fixed it in the next step). Such mistakes
are caught by looking at the theorem statement and at \lean{\#print axioms},
not by reading hundreds of thousands of lines of proof. Our checklist at each milestone was short:
no \lean{sorry}, no \lean{native\_decide}, the expected axioms only, and an
unchanged statement.

\paragraph{Specifications are the unit of parallelism.}
Because every function's proof depends only on the specifications of its callees
(\cref{sec:proof}), proofs can be done in any order and by different
agents. In the first week of the x86-64 work, a single agent built the model,
the bridge lemmas and the shared infrastructure, and verified the first few
functions. Once the specifications were in place and the work was
parallelized, the remaining functions, including the JIT, took three days.
The coordinator's job was then mostly to merge branches, rebuild, and pass
fixes from one worker to another (for example, a workaround for a proof that
the elaborator accepted but the kernel rejected).

\paragraph{Mechanical size is cheap.}
The proof is about 75 times the size of the code. For humans this would be
prohibitive: seL4's verification took about 20 person-years of effort, of
which about 11 were for the seL4-specific proof and the rest for frameworks
and tools~\cite{sel4}. xv6iris, also built with agents, took 102 calendar days
including its framework~\cite{xv6iris}. (Person-years and calendar days are
different units; \safejit took 29 calendar days.) For agents, the size of a proof matters much less
than its difficulty, and much of \safejit's proof is easy but long:
symbolic execution of straight-line code, one lemma per instruction form,
one template lemma per opcode. Investments that make proofs more
mechanical (scripts that generate bridge lemmas, tactics like \lean{twalk},
symbolic addresses that let proofs survive a relink) paid off more than
investments that make them shorter.

\paragraph{Executable models catch mistakes early.}
The JIT model is a Lean function that can run, and testing it against the C
code on every compilation in the test suite (\cref{sec:eval}) found most
mistakes in the model in seconds, long before a proof attempt
would have.

\paragraph{The trusted base needs adversarial review.}
The agent that wrote the proofs also wrote the process model, and both bugs
of \cref{sec:findings} hid in that model. A model that the proof's author
chose is a weak point whether the author is a human or an agent. Asking a
different agent to attack the claim was cheap and effective, and we would
make it a routine step.

\paragraph{Resources.}
Build time was a constant concern. The Lean build grew to over 1{,}400 jobs,
and some single proofs took tens of CPU-minutes to check. The agents spent
noticeable effort on build performance (splitting modules, reducing imports,
replacing expensive tactic calls), often at the researcher's prompting. With
several agents building at once, the machine (128 cores) ran into its
file-descriptor limit.

\section{Discussion}
\label{sec:discussion}

\subsection{Binary verification does not make compilers obsolete}

Kaashoek and Zeldovich argue that verifying the binary ``allows MachCSL to
factor out the compiler, linker, assembler, and other parts of the build process
from the trusted computing base''~\cite{xv6iris}, and that the proofs absorb
gcc's optimizations for free. \safejit relies on the same idea: gcc is
untrusted, and we never reason about C.

A JIT shows where this argument stops. The machine code a JIT runs does not
exist when the proof is done. It is generated later, from programs and type
feedback chosen by the attacker. The binary can be verified, but one function in
it is a compiler, and its specification must quantify over every program
it could be given and every code it could emit. ``For every well-formed
function and every feedback vector, the emitted code preserves the
invariants'' is a compiler-verification theorem, and most of the
work in \safejit's proof is spent there: template lemmas for each opcode,
soundness of the abstract types that remove guards, and the consistency of deopt and OSR
exits with the interpreter. Binary-level verification removes gcc from the
trusted base; it does not remove the need to verify the compilers that are
part of the system. JIT engines are such systems, and they are some of the most
attacked software in existence. Verified-compiler research is still needed, now
inside binary-level proofs.

The two-layer structure of the JIT proof (\cref{sec:jit}) is how we combine
them: a model of the JIT that is a Lean function, proved safe with the usual
tools of compiler verification, and a proof that the binary of
\code{jit\_compile} refines that model. The refinement layer is mechanical
(about 12k lines for the RISC-V JIT) and does not grow with the
sophistication of the analysis, only with the size of the code generator.

\subsection{Limitations}

\paragraph{Safety, not correctness.}
The theorem says that the engine never gets stuck. It does not say
that it computes the right answer, that it terminates, or that it is free of
side channels. Speculative-execution attacks are outside the model, as they
are for most verified systems; V8's developers concluded that code-level
mitigations cannot fully contain Spectre and turned to process
isolation~\cite{spectre-here-to-stay}. A wrong answer
from the JIT is still a serious bug, but it is not an exploitable one in the
sense of \cref{sec:intro}: in a production engine, type confusion is dangerous
because it leads to memory corruption, and memory corruption is what the theorem rules out.

\paragraph{A small engine.}
\safejit lacks closures, exceptions, \code{eval}, most of the standard library,
regular expressions, typed arrays and concurrency. Its JIT is a baseline JIT
with simple register allocation; it has no optimizing tier. We believe that
the missing language features are ``more of the same'': each new heap kind
extends the heap invariant and the interpreter's opcode proofs, and JIT fast
paths for them have the same shape as the existing paths for arrays and objects (tag guard,
kind guard, bounds or shape check, then a direct access). Native library
functions are proved separately against a fixed interface and do not touch
the JIT proof. An optimizing tier is a different matter. It is where most
production JIT bugs are, and verifying one is verified-compiler work at scale.
The structure of our proof carries over, but the effort will be larger.

\paragraph{Memory reclamation.}
The collector frees strings, arrays and objects, but not functions or JIT code.
Functions are immortal anyway, and since a function is compiled at most
$N$ times (the \code{--max-deopts} limit, 8 by default), the code leaked by
deopts is at most $N$ times a constant multiple of the program's size. Reclaiming code becomes necessary once functions can die,
and it is a cross-cutting change: code ownership (\lean{codeAt}) is a
persistent fact in the proof, so a running activation would need
non-persistent evidence that its code stays mapped. We plan to keep the live
code in a per-epoch invariant that the collector opens, and to count the
activations of each piece of code.

\paragraph{Single-threaded.}
There is no background compilation and no concurrent collection. The logic
supports concurrency (MachCSL verifies multi-core xv6), but the engine would
need the most changes.

\paragraph{Trusted base.}
The theorem trusts the Sail models, our process model of Linux, Lean's kernel,
and the LRAT checker of \lean{bv_decide} (\cref{sec:theorem}). The process
model is ours, and \cref{sec:findings} shows that it can be wrong in
ways that matter. On x86-64, the floating-point instructions are opaque, and the proof
assumes that the MXCSR register keeps its default value (\cref{sec:x86}).

\subsection{Future work}

In order of value, the next steps are:
\begin{enumerate}
\item \textbf{Bounds-check elimination.} Array accesses always check the index.
  Eliminating the check when the index is known to be in range, such as a loop
  counter compared against \code{length}, needs range facts on top of the loop type
  facts, and an invariant that the array does not shrink in the meantime. Bugs
  in this area are subtle: in CVE-2019-8518, loop-invariant code motion
  hoisted an array access ahead of the bounds check that guarded
  it~\cite{cve-2019-8518}.
\item \textbf{Wider register allocation.} Only five locals live in registers,
  and pending stack slots are flushed at every merge point. Keeping pending state
  across merges would remove most of the remaining frame traffic in loops.
\item \textbf{Parity for x86-64.} The x86-64 JIT has no register allocation,
  virtual stack or loop facts. Porting them is mostly a matter of porting the
  RISC-V model and proofs, which is the kind of work that agents do well.
\item \textbf{Code reclamation and closures}, as described above.
\item \textbf{eBPF.} An eBPF verifier and JIT in the same framework would be a
  second demonstrator with immediate practical relevance, and the verifier's range
  analysis is the kind of analysis that bounds-check elimination needs.
\end{enumerate}

\section{Related Work}
\label{sec:related}

\Cref{tab:related} compares \safejit with the closest prior work.

\begin{table*}
\caption{Verified JITs and related systems. ``Speculation'' means guards on
dynamically observed types with deoptimization to a lower tier.}
\label{tab:related}
\small
\begin{tabular}{@{}llllc@{}}
\toprule
Work & What is verified & Level & Theorem & Speculation \\
\midrule
Myreen~\cite{myreen-jit} & JIT for a small bytecode language & x86 machine code & functional correctness & no \\
Jitawa~\cite{jitawa} & Lisp runtime, GC and JIT & x86-64 machine code & functional correctness & no \\
Jitk~\cite{jitk} & classic BPF JIT & Coq, extracted with CompCert & functional correctness & no \\
Jitterbug~\cite{jitterbug} & Linux eBPF JITs, per instruction & Rosette model of the C & functional correctness & no \\
CertrBPF~\cite{certrbpf} & eBPF VM for RIOT & Coq, C & correctness, isolation & no \\
Agni~\cite{agni} & Linux eBPF verifier range analysis & SMT & soundness & -- \\
VeRA~\cite{vera} & Firefox JIT range analysis & C++ pass, SMT & soundness of one analysis & -- \\
Icarus~\cite{icarus} & CacheIR stub generators & macro-assembler model & safety of stubs & IC guards \\
sourir~\cite{sourir} & speculative optimizations & IR & semantic preservation & yes \\
CoreJIT~\cite{corejit,corejit-native} & JIT with speculation and deopt & Coq model, CompCert back end & semantic preservation & yes \\
VeriWasm~\cite{veriwasm} & Wasm AOT output & x86 binary checker & sandbox safety & no \\
vWasm~\cite{vwasm} & Wasm compiler & F$^\star$ & sandbox safety & no \\
Crocus~\cite{crocus} & Cranelift lowering rules & SMT & rule correctness & no \\
CakeML~\cite{cakeml,cakeml-backend,cakeml-eval} & ML compiler, runtime and \code{Eval} & machine code & functional correctness & no \\
Islaris~\cite{islaris} & machine-code fragments & Sail Arm/RISC-V & functional correctness & -- \\
xv6iris~\cite{xv6iris} & xv6 kernel and user programs & RISC-V binary, Sail & functional correctness & -- \\
\midrule
\safejit & whole engine incl.\ JIT output & RISC-V and x86-64 binaries, Sail & safety for all inputs & yes, plus OSR \\
\bottomrule
\end{tabular}
\end{table*}

\paragraph{Speculation and deoptimization.}
Speculation with deoptimization goes back to Smalltalk and
Self~\cite{smalltalk84,self89,deopt92}. Guo and Palsberg~\cite{traces}
formalized trace-based compilation, including hot-path recording and side
exits, with a bisimulation criterion. Flückiger et al.~\cite{sourir} gave a
formal account of speculation with explicit assumptions and deoptimization
across program versions: sourir is an IR with \code{assume} instructions that
deoptimize to a less optimized version, and they prove a set of optimizations
correct in their presence. D'Elia and Demetrescu~\cite{osr-distilled} give a
correctness argument for OSR that reconstructs the state across program
versions. Preserving a program's live values across a transition is stronger
than what we prove (that the destination frame is safe to resume), but it is
not a theorem about an engine binary.
CoreJIT~\cite{corejit} is a JIT verified in Coq that speculates, deoptimizes and
performs dynamic optimization over a CompCert-style IR. It proves semantic
preservation of the whole dynamic compilation process. Its
successor~\cite{corejit-native} adds a native back end through CompCert and
reasons about the transitions between interpreted and generated code down to
CompCert's x86 assembly semantics; it trusts the step from assembly syntax to
installed machine bytes and the implementations of runtime primitives. These
are correctness proofs, and stronger than ours in that sense. \safejit's
theorem is weaker (safety), but its trusted boundary is the ISA model: the
interpreter, runtime, collector, libc and the bytes of the generated code are
all part of the verified binary.
Our JIT's abstract types, loop facts and deopt metadata are much simpler than
the optimizations these works handle.

\paragraph{Verified JITs for real code.}
Myreen~\cite{myreen-jit} verified a JIT for a small stack-based bytecode language
down to x86 machine code in HOL4, including self-modifying code, using his
decompilation into logic. Jitk~\cite{jitk} verified a JIT for classic BPF
using CompCert, and Jitterbug~\cite{jitterbug} verified the Linux kernel's
eBPF JITs instruction by instruction, finding many bugs. CertrBPF~\cite{certrbpf} is
a verified eBPF interpreter for an embedded OS, and Agni~\cite{agni} and
VeRA~\cite{vera} verify range analyses in the Linux eBPF verifier and in
Firefox's JIT, respectively. TurboTV~\cite{turbotv} validates individual
TurboFan IR transformations in V8 with SMT; a check of one translation is
neither a test oracle nor a proof about all emitted programs. These systems
target languages without dynamic types, or verify one component, and none of
them speculates with deoptimization.

Jitawa~\cite{jitawa} is the closest precedent for a dynamic-language runtime
verified together with its JIT at the machine-code level: its parser, printer,
copying collector and JIT compiler for a Lisp were verified in HOL4 down to
x86-64 code, with a functional-correctness theorem whose premises include the
initial memory and the execution of the source program. It compiles each
function before running it, without feedback, guards, deoptimization or OSR.
Icarus~\cite{icarus} is the closest precedent for type-specialized fast paths
in a real JavaScript engine. It verifies safety properties of the inline-cache
stubs that SpiderMonkey's CacheIR generators can emit, by symbolic
meta-execution over models of the macro-assembler, and integrates the
generated code back into Firefox. Its unit of verification is the stub
generator and the modeled operations; ours is the engine image, the emitted
instruction bytes and the transitions between tiers. An inline-cache miss,
which falls back to a generic path, is also simpler than a whole-frame
deoptimization.

\paragraph{WebAssembly.}
For WebAssembly, VeriWasm~\cite{veriwasm} checks that the x86 output of an
ahead-of-time compiler stays in its sandbox, vWasm~\cite{vwasm} is a
compiler verified in F$^\star$ to produce sandboxed code, and
Crocus~\cite{crocus} verifies the instruction-lowering rules of Cranelift.
Like \safejit, VeriWasm and vWasm prove a safety property rather than
correctness, which is the property that matters for a sandbox. None of these
verifies speculation; production engines do speculate on WebAssembly, for
example V8 inlines indirect calls based on feedback and deoptimizes when the
assumption fails~\cite{v8-wasm-spec}.

\paragraph{Machine-code verification.}
CompCert~\cite{compcert} and CakeML~\cite{cakeml,cakeml-backend} are verified
compilers; CakeML's proofs go down to machine code and include its runtime
and garbage collector. CakeML also compiles at run time: its verified
interactive REPL~\cite{cakeml} and its verified \code{Eval}~\cite{cakeml-eval}
compile and install new code that exchanges values with existing code, so
run-time code generation alone does not distinguish \safejit; feedback-directed
speculation does. Cai, Shao and Vaynberg~\cite{cai-smc} gave a Coq framework
for local reasoning about code that is stored and generated in writable
memory. Our \lean{codeAt} resource applies the same idea to concrete engine
code, with page permissions and the RISC-V cache-flush contract. Islaris~\cite{islaris} introduced separation logic over
Sail ISA models, using Isla to compute instruction traces and Iris for
reasoning. xv6iris~\cite{xv6iris} applied the idea to a whole system: it
verifies the RISC-V binary of the xv6 kernel and user programs against the Sail
model in Lean, with a concurrent separation logic, MachCSL, built on
iris-lean~\cite{iris,iris-lean}. Its proofs were also developed with AI
agents, and found 11 bugs in xv6. \safejit builds directly on xv6iris: we
reuse its Sail model configuration, its fork of the Sail Lean backend, and
parts of MachCSL (access kinds, the boot sequence, and board writes). What is new
in \safejit is what a JIT needs on top of this: code as a resource that
is created at run time and installed through \code{mprotect} and a cache flush,
a JIT whose output is verified for every input, and the proof structure of
\cref{sec:proof-compiler,sec:jit} that connects the two. seL4~\cite{sel4}
remains the reference point for the effort that a full functional-correctness
proof of a systems component takes.

\paragraph{Testing JITs.}
Fuzzers such as Fuzzilli~\cite{fuzzilli} and JIT-Picking~\cite{jitpicking}
find many JIT bugs, and differential testing of a JIT against its own
interpreter is a standard oracle. We used the same techniques during
development (\cref{sec:eval}). They find bugs; a proof shows their absence.

\section{Conclusion}
\label{sec:conclusion}

\safejit is a JavaScript engine with a speculating JIT compiler whose RISC-V and
x86-64 executables are proved, in Lean against the Sail ISA models, never to
get stuck on any input. The proof covers everything that runs: the parser and
bytecode compiler, the interpreter, the runtime and garbage collector, the
freestanding libc, the JIT compiler, and every piece of code that the JIT
could emit for any program and any type feedback, including its guards,
deoptimization exits and on-stack-replacement entries. Neither the C compiler
nor floating-point arithmetic is trusted.

The engine is small, and its JIT is a baseline JIT. But the parts that make
JITs dangerous, speculation over attacker-controlled feedback, removal of guards
based on abstract types, and transitions between tiers, are all there, and
they are verified at the level where an attacker operates. Two lessons stand
out. First, binary-level verification does not remove the need for compiler
verification when the binary contains a compiler; the two have to be combined,
and a model of the JIT that is proved safe and refined by the binary is an
effective way to combine them. Second, AI agents change the economics of this
kind of work. The proof is 75 times the size of the code and took four weeks,
with one researcher whose role was to set goals, steer the design, and check
the statement and the trusted base. We think this makes it plausible that
production JIT engines, and other security-critical systems that were long
considered out of reach, can be verified at the level of their binaries.

\begin{acks}
\safejit builds on xv6iris by M.\ Frans Kaashoek and Nickolai Zeldovich: its
Sail model configuration and Lean backend fork, and parts of its MachCSL
framework (MIT license). We thank the authors of Sail, the Sail RISC-V and
x86 models, Lean, and iris-lean, on which this work depends.
\end{acks}

\bibliographystyle{ACM-Reference-Format}
\bibliography{refs}


\begin{thebibliography}{45}

%%% the Web address.

\ifx \showCODEN    \undefined \def \showCODEN     #1{\unskip}     \fi
\ifx \showDOI      \undefined \def \showDOI       #1{#1}\fi
\ifx \showISBNx    \undefined \def \showISBNx     #1{\unskip}     \fi
\ifx \showISBNxiii \undefined \def \showISBNxiii  #1{\unskip}     \fi
\ifx \showISSN     \undefined \def \showISSN      #1{\unskip}     \fi
\ifx \showLCCN     \undefined \def \showLCCN      #1{\unskip}     \fi
\ifx \shownote     \undefined \def \shownote      #1{#1}          \fi
\ifx \showarticletitle \undefined \def \showarticletitle #1{#1}   \fi
\ifx \showURL      \undefined \def \showURL       {\relax}        \fi
% invisible to TeX
\providecommand\bibfield[2]{#2}
\providecommand\bibinfo[2]{#2}
\providecommand\natexlab[1]{#1}
\providecommand\showeprint[2][]{arXiv:#2}

\bibitem[{Apple}(2024)]%
        {lockdown}
\bibfield{author}{\bibinfo{person}{{Apple}}.} \bibinfo{year}{2024}\natexlab{}.
\newblock \bibinfo{title}{Harden Your Devices with {Lockdown Mode}}.
\newblock \bibinfo{howpublished}{Apple Personal Safety User Guide}.
\newblock
\urldef\tempurl%
\url{https://support.apple.com/guide/personal-safety/harden-your-devices-with-lockdown-mode-ipsd5baf79d0/web}
\showURL{%
\tempurl}


\bibitem[Armstrong et~al\mbox{.}(2019)]%
        {sail}
\bibfield{author}{\bibinfo{person}{Alasdair Armstrong}, \bibinfo{person}{Thomas
  Bauereiss}, \bibinfo{person}{Brian Campbell}, \bibinfo{person}{Alastair
  Reid}, \bibinfo{person}{Kathryn~E. Gray}, \bibinfo{person}{Robert~M. Norton},
  \bibinfo{person}{Prashanth Mundkur}, \bibinfo{person}{Mark Wassell},
  \bibinfo{person}{Jon French}, \bibinfo{person}{Christopher Pulte},
  \bibinfo{person}{Shaked Flur}, \bibinfo{person}{Ian Stark},
  \bibinfo{person}{Neel Krishnaswami}, {and} \bibinfo{person}{Peter Sewell}.}
  \bibinfo{year}{2019}\natexlab{}.
\newblock \showarticletitle{{ISA} Semantics for {ARMv8-A}, {RISC-V}, and
  {CHERI-MIPS}}.
\newblock \bibinfo{journal}{\emph{Proc.\ ACM Program.\ Lang.}}
  \bibinfo{volume}{3}, \bibinfo{number}{POPL} (\bibinfo{year}{2019}).
\newblock
\urldef\tempurl%
\url{https://doi.org/10.1145/3290384}
\showDOI{\tempurl}


\bibitem[Barri{\`e}re et~al\mbox{.}(2021)]%
        {corejit}
\bibfield{author}{\bibinfo{person}{Aur{\`e}le Barri{\`e}re},
  \bibinfo{person}{Sandrine Blazy}, \bibinfo{person}{Olivier Fl{\"u}ckiger},
  \bibinfo{person}{David Pichardie}, {and} \bibinfo{person}{Jan Vitek}.}
  \bibinfo{year}{2021}\natexlab{}.
\newblock \showarticletitle{Formally Verified Speculation and Deoptimization in
  a {JIT} Compiler}.
\newblock \bibinfo{journal}{\emph{Proc.\ ACM Program.\ Lang.}}
  \bibinfo{volume}{5}, \bibinfo{number}{POPL} (\bibinfo{year}{2021}).
\newblock
\urldef\tempurl%
\url{https://doi.org/10.1145/3434327}
\showDOI{\tempurl}


\bibitem[Barri{\`e}re et~al\mbox{.}(2023)]%
        {corejit-native}
\bibfield{author}{\bibinfo{person}{Aur{\`e}le Barri{\`e}re},
  \bibinfo{person}{Sandrine Blazy}, {and} \bibinfo{person}{David Pichardie}.}
  \bibinfo{year}{2023}\natexlab{}.
\newblock \showarticletitle{Formally Verified Native Code Generation in an
  Effectful {JIT}: Turning the {CompCert} Backend into a Formally Verified
  {JIT} Compiler}.
\newblock \bibinfo{journal}{\emph{Proc.\ ACM Program.\ Lang.}}
  \bibinfo{volume}{7}, \bibinfo{number}{POPL} (\bibinfo{year}{2023}).
\newblock
\urldef\tempurl%
\url{https://doi.org/10.1145/3571202}
\showDOI{\tempurl}


\bibitem[Bernhard et~al\mbox{.}(2022)]%
        {jitpicking}
\bibfield{author}{\bibinfo{person}{Lukas Bernhard}, \bibinfo{person}{Tobias
  Scharnowski}, \bibinfo{person}{Moritz Schloegel}, \bibinfo{person}{Tim
  Blazytko}, {and} \bibinfo{person}{Thorsten Holz}.}
  \bibinfo{year}{2022}\natexlab{}.
\newblock \showarticletitle{{JIT-Picking}: Differential Fuzzing of {JavaScript}
  Engines}. In \bibinfo{booktitle}{\emph{ACM SIGSAC Conference on Computer and
  Communications Security (CCS)}}. \bibinfo{pages}{351--364}.
\newblock
\urldef\tempurl%
\url{https://doi.org/10.1145/3548606.3560624}
\showDOI{\tempurl}


\bibitem[Bosamiya et~al\mbox{.}(2022)]%
        {vwasm}
\bibfield{author}{\bibinfo{person}{Jay Bosamiya}, \bibinfo{person}{Wen~Shih
  Lim}, {and} \bibinfo{person}{Bryan Parno}.} \bibinfo{year}{2022}\natexlab{}.
\newblock \showarticletitle{Provably-Safe Multilingual Software Sandboxing
  using {WebAssembly}}. In \bibinfo{booktitle}{\emph{USENIX Security
  Symposium}}. \bibinfo{pages}{1975--1992}.
\newblock
\urldef\tempurl%
\url{https://www.usenix.org/conference/usenixsecurity22/presentation/bosamiya}
\showURL{%
\tempurl}


\bibitem[Brown et~al\mbox{.}(2020)]%
        {vera}
\bibfield{author}{\bibinfo{person}{Fraser Brown}, \bibinfo{person}{John
  Renner}, \bibinfo{person}{Andres N{\"o}tzli}, \bibinfo{person}{Sorin Lerner},
  \bibinfo{person}{Hovav Shacham}, {and} \bibinfo{person}{Deian Stefan}.}
  \bibinfo{year}{2020}\natexlab{}.
\newblock \showarticletitle{Towards a Verified Range Analysis for {JavaScript}
  {JITs}}. In \bibinfo{booktitle}{\emph{ACM SIGPLAN Conference on Programming
  Language Design and Implementation (PLDI)}}. \bibinfo{pages}{135--150}.
\newblock
\urldef\tempurl%
\url{https://doi.org/10.1145/3385412.3385968}
\showDOI{\tempurl}


\bibitem[Cai et~al\mbox{.}(2007)]%
        {cai-smc}
\bibfield{author}{\bibinfo{person}{Hongxu Cai}, \bibinfo{person}{Zhong Shao},
  {and} \bibinfo{person}{Alexander Vaynberg}.} \bibinfo{year}{2007}\natexlab{}.
\newblock \showarticletitle{Certified Self-Modifying Code}. In
  \bibinfo{booktitle}{\emph{ACM SIGPLAN Conference on Programming Language
  Design and Implementation (PLDI)}}. \bibinfo{pages}{66--77}.
\newblock
\urldef\tempurl%
\url{https://doi.org/10.1145/1250734.1250743}
\showDOI{\tempurl}


\bibitem[Chambers et~al\mbox{.}(1989)]%
        {self89}
\bibfield{author}{\bibinfo{person}{Craig Chambers}, \bibinfo{person}{David
  Ungar}, {and} \bibinfo{person}{Elgin Lee}.} \bibinfo{year}{1989}\natexlab{}.
\newblock \showarticletitle{An Efficient Implementation of {SELF}, a
  Dynamically-Typed Object-Oriented Language Based on Prototypes}. In
  \bibinfo{booktitle}{\emph{Conference on Object-Oriented Programming Systems,
  Languages and Applications (OOPSLA)}}. \bibinfo{pages}{49--70}.
\newblock
\urldef\tempurl%
\url{https://doi.org/10.1145/74877.74884}
\showDOI{\tempurl}


\bibitem[de~Moura and Ullrich(2021)]%
        {lean4}
\bibfield{author}{\bibinfo{person}{Leonardo de Moura} {and}
  \bibinfo{person}{Sebastian Ullrich}.} \bibinfo{year}{2021}\natexlab{}.
\newblock \showarticletitle{The {Lean} 4 Theorem Prover and Programming
  Language}. In \bibinfo{booktitle}{\emph{Automated Deduction -- CADE 28}}
  \emph{(\bibinfo{series}{LNCS}, Vol.~\bibinfo{volume}{12699})}.
  \bibinfo{pages}{625--635}.
\newblock
\urldef\tempurl%
\url{https://doi.org/10.1007/978-3-030-79876-5_37}
\showDOI{\tempurl}


\bibitem[D'Elia and Demetrescu(2018)]%
        {osr-distilled}
\bibfield{author}{\bibinfo{person}{Daniele~Cono D'Elia} {and}
  \bibinfo{person}{Camil Demetrescu}.} \bibinfo{year}{2018}\natexlab{}.
\newblock \showarticletitle{On-Stack Replacement, Distilled}. In
  \bibinfo{booktitle}{\emph{ACM SIGPLAN Conference on Programming Language
  Design and Implementation (PLDI)}}. \bibinfo{pages}{166--180}.
\newblock
\urldef\tempurl%
\url{https://doi.org/10.1145/3192366.3192396}
\showDOI{\tempurl}


\bibitem[Deutsch and Schiffman(1984)]%
        {smalltalk84}
\bibfield{author}{\bibinfo{person}{L.~Peter Deutsch} {and}
  \bibinfo{person}{Allan~M. Schiffman}.} \bibinfo{year}{1984}\natexlab{}.
\newblock \showarticletitle{Efficient Implementation of the {Smalltalk-80}
  System}. In \bibinfo{booktitle}{\emph{ACM SIGACT-SIGPLAN Symposium on
  Principles of Programming Languages (POPL)}}. \bibinfo{pages}{297--302}.
\newblock
\urldef\tempurl%
\url{https://doi.org/10.1145/800017.800542}
\showDOI{\tempurl}


\bibitem[Fl{\"u}ckiger et~al\mbox{.}(2018)]%
        {sourir}
\bibfield{author}{\bibinfo{person}{Olivier Fl{\"u}ckiger},
  \bibinfo{person}{Gabriel Scherer}, \bibinfo{person}{Ming-Ho Yee},
  \bibinfo{person}{Aviral Goel}, \bibinfo{person}{Amal Ahmed}, {and}
  \bibinfo{person}{Jan Vitek}.} \bibinfo{year}{2018}\natexlab{}.
\newblock \showarticletitle{Correctness of Speculative Optimizations with
  Dynamic Deoptimization}.
\newblock \bibinfo{journal}{\emph{Proc.\ ACM Program.\ Lang.}}
  \bibinfo{volume}{2}, \bibinfo{number}{POPL} (\bibinfo{year}{2018}),
  \bibinfo{pages}{49:1--49:28}.
\newblock
\urldef\tempurl%
\url{https://doi.org/10.1145/3158137}
\showDOI{\tempurl}


\bibitem[{Google Threat Analysis Group and Mandiant}(2024)]%
        {tag2024}
\bibfield{author}{\bibinfo{person}{{Google Threat Analysis Group and
  Mandiant}}.} \bibinfo{year}{2024}\natexlab{}.
\newblock \bibinfo{title}{We're All in this Together: A Year in Review of
  Zero-Days Exploited In-the-Wild in 2023}.
\newblock
\newblock
\urldef\tempurl%
\url{https://storage.googleapis.com/gweb-uniblog-publish-prod/documents/Year_in_Review_of_ZeroDays.pdf}
\showURL{%
\tempurl}


\bibitem[Gro{\ss}(2019)]%
        {cve-2019-8518}
\bibfield{author}{\bibinfo{person}{Samuel Gro{\ss}}.}
  \bibinfo{year}{2019}\natexlab{}.
\newblock \bibinfo{title}{{JavaScriptCore}: {OOB} Access in {FTL} {JIT} due to
  {LICM} Moving Array Access before the Bounds Check}.
\newblock \bibinfo{howpublished}{Project Zero issue 42450850 (CVE-2019-8518)}.
\newblock
\urldef\tempurl%
\url{https://project-zero.issues.chromium.org/issues/42450850}
\showURL{%
\tempurl}


\bibitem[Gro{\ss}(2024)]%
        {v8sandbox}
\bibfield{author}{\bibinfo{person}{Samuel Gro{\ss}}.}
  \bibinfo{year}{2024}\natexlab{}.
\newblock \bibinfo{title}{The {V8} Sandbox}.
\newblock \bibinfo{howpublished}{V8 blog}.
\newblock
\urldef\tempurl%
\url{https://v8.dev/blog/sandbox}
\showURL{%
\tempurl}


\bibitem[Gro{\ss} et~al\mbox{.}(2023)]%
        {fuzzilli}
\bibfield{author}{\bibinfo{person}{Samuel Gro{\ss}}, \bibinfo{person}{Simon
  Koch}, \bibinfo{person}{Lukas Bernhard}, \bibinfo{person}{Thorsten Holz},
  {and} \bibinfo{person}{Martin Johns}.} \bibinfo{year}{2023}\natexlab{}.
\newblock \showarticletitle{{FUZZILLI}: Fuzzing for {JavaScript} {JIT} Compiler
  Vulnerabilities}. In \bibinfo{booktitle}{\emph{Network and Distributed System
  Security Symposium (NDSS)}}.
\newblock
\urldef\tempurl%
\url{https://doi.org/10.14722/ndss.2023.24290}
\showDOI{\tempurl}


\bibitem[H{\"o}lzle et~al\mbox{.}(1992)]%
        {deopt92}
\bibfield{author}{\bibinfo{person}{Urs H{\"o}lzle}, \bibinfo{person}{Craig
  Chambers}, {and} \bibinfo{person}{David Ungar}.}
  \bibinfo{year}{1992}\natexlab{}.
\newblock \showarticletitle{Debugging Optimized Code with Dynamic
  Deoptimization}. In \bibinfo{booktitle}{\emph{ACM SIGPLAN Conference on
  Programming Language Design and Implementation (PLDI)}}.
  \bibinfo{pages}{32--43}.
\newblock
\urldef\tempurl%
\url{https://doi.org/10.1145/143095.143114}
\showDOI{\tempurl}


\bibitem[Johnson et~al\mbox{.}(2021)]%
        {veriwasm}
\bibfield{author}{\bibinfo{person}{Evan Johnson}, \bibinfo{person}{David
  Thien}, \bibinfo{person}{Yousef Alhessi}, \bibinfo{person}{Shravan Narayan},
  \bibinfo{person}{Fraser Brown}, \bibinfo{person}{Sorin Lerner},
  \bibinfo{person}{Tyler McMullen}, \bibinfo{person}{Stefan Savage}, {and}
  \bibinfo{person}{Deian Stefan}.} \bibinfo{year}{2021}\natexlab{}.
\newblock \showarticletitle{{\emph{Doveryai, no Proveryai}}: {SFI} Safety for
  Native-Compiled {Wasm}}. In \bibinfo{booktitle}{\emph{Network and Distributed
  System Security Symposium (NDSS)}}.
\newblock
\urldef\tempurl%
\url{https://doi.org/10.14722/ndss.2021.24078}
\showDOI{\tempurl}


\bibitem[Jung et~al\mbox{.}(2018)]%
        {iris}
\bibfield{author}{\bibinfo{person}{Ralf Jung}, \bibinfo{person}{Robbert
  Krebbers}, \bibinfo{person}{Jacques-Henri Jourdan}, \bibinfo{person}{Ale{\v
  s} Bizjak}, \bibinfo{person}{Lars Birkedal}, {and} \bibinfo{person}{Derek
  Dreyer}.} \bibinfo{year}{2018}\natexlab{}.
\newblock \showarticletitle{Iris from the Ground Up: A Modular Foundation for
  Higher-Order Concurrent Separation Logic}.
\newblock \bibinfo{journal}{\emph{Journal of Functional Programming}}
  \bibinfo{volume}{28} (\bibinfo{year}{2018}), \bibinfo{pages}{e20}.
\newblock
\urldef\tempurl%
\url{https://doi.org/10.1017/S0956796818000151}
\showDOI{\tempurl}


\bibitem[Kaashoek and Zeldovich(2026)]%
        {xv6iris}
\bibfield{author}{\bibinfo{person}{M.~Frans Kaashoek} {and}
  \bibinfo{person}{Nickolai Zeldovich}.} \bibinfo{year}{2026}\natexlab{}.
\newblock \bibinfo{title}{Extending Concurrent Separation Logic to the Hardware
  Level to Verify the xv6 {OS} Kernel on {RISC-V} with {AI} Agents}.
\newblock
\newblock
\urldef\tempurl%
\url{https://doi.org/10.48550/arXiv.2609.04043}
\showDOI{\tempurl}
\showeprint[arxiv]{2609.04043v3}~[cs.LO]
\newblock
\shownote{Version 3}.


\bibitem[Klein et~al\mbox{.}(2009)]%
        {sel4}
\bibfield{author}{\bibinfo{person}{Gerwin Klein}, \bibinfo{person}{Kevin
  Elphinstone}, \bibinfo{person}{Gernot Heiser}, \bibinfo{person}{June
  Andronick}, \bibinfo{person}{David Cock}, \bibinfo{person}{Philip Derrin},
  \bibinfo{person}{Dhammika Elkaduwe}, \bibinfo{person}{Kai Engelhardt},
  \bibinfo{person}{Rafal Kolanski}, \bibinfo{person}{Michael Norrish},
  \bibinfo{person}{Thomas Sewell}, \bibinfo{person}{Harvey Tuch}, {and}
  \bibinfo{person}{Simon Winwood}.} \bibinfo{year}{2009}\natexlab{}.
\newblock \showarticletitle{{seL4}: Formal Verification of an {OS} Kernel}. In
  \bibinfo{booktitle}{\emph{ACM Symposium on Operating Systems Principles
  (SOSP)}}. \bibinfo{pages}{207--220}.
\newblock
\urldef\tempurl%
\url{https://doi.org/10.1145/1629575.1629596}
\showDOI{\tempurl}


\bibitem[Kumar et~al\mbox{.}(2014)]%
        {cakeml}
\bibfield{author}{\bibinfo{person}{Ramana Kumar}, \bibinfo{person}{Magnus~O.
  Myreen}, \bibinfo{person}{Michael Norrish}, {and} \bibinfo{person}{Scott
  Owens}.} \bibinfo{year}{2014}\natexlab{}.
\newblock \showarticletitle{{CakeML}: A Verified Implementation of {ML}}. In
  \bibinfo{booktitle}{\emph{ACM SIGPLAN-SIGACT Symposium on Principles of
  Programming Languages (POPL)}}. \bibinfo{pages}{179--191}.
\newblock
\urldef\tempurl%
\url{https://doi.org/10.1145/2535838.2535841}
\showDOI{\tempurl}


\bibitem[Kwon et~al\mbox{.}(2024)]%
        {turbotv}
\bibfield{author}{\bibinfo{person}{Seungwan Kwon}, \bibinfo{person}{Jaeseong
  Kwon}, \bibinfo{person}{Wooseok Kang}, \bibinfo{person}{Juneyoung Lee}, {and}
  \bibinfo{person}{Kihong Heo}.} \bibinfo{year}{2024}\natexlab{}.
\newblock \showarticletitle{Translation Validation for {JIT} Compiler in the
  {V8} {JavaScript} Engine}. In \bibinfo{booktitle}{\emph{IEEE/ACM
  International Conference on Software Engineering (ICSE)}}.
  \bibinfo{pages}{1--12}.
\newblock
\urldef\tempurl%
\url{https://doi.org/10.1145/3597503.3639189}
\showDOI{\tempurl}


\bibitem[{Lean community}(2026)]%
        {iris-lean}
\bibfield{author}{\bibinfo{person}{{Lean community}}.}
  \bibinfo{year}{2026}\natexlab{}.
\newblock \bibinfo{title}{iris-lean: {Iris} in {Lean} 4}.
\newblock \bibinfo{howpublished}{GitHub repository}.
\newblock
\urldef\tempurl%
\url{https://github.com/leanprover-community/iris-lean}
\showURL{%
\tempurl}


\bibitem[Lehmann and Liedtke(2025)]%
        {v8-wasm-spec}
\bibfield{author}{\bibinfo{person}{Daniel Lehmann} {and}
  \bibinfo{person}{Matthias Liedtke}.} \bibinfo{year}{2025}\natexlab{}.
\newblock \bibinfo{title}{Speculative Optimizations for {WebAssembly} using
  Deopts and Inlining}.
\newblock \bibinfo{howpublished}{V8 blog}.
\newblock
\urldef\tempurl%
\url{https://v8.dev/blog/wasm-speculative-optimizations}
\showURL{%
\tempurl}


\bibitem[Leroy(2009)]%
        {compcert}
\bibfield{author}{\bibinfo{person}{Xavier Leroy}.}
  \bibinfo{year}{2009}\natexlab{}.
\newblock \showarticletitle{Formal Verification of a Realistic Compiler}.
\newblock \bibinfo{journal}{\emph{Commun. ACM}} \bibinfo{volume}{52},
  \bibinfo{number}{7} (\bibinfo{year}{2009}), \bibinfo{pages}{107--115}.
\newblock
\urldef\tempurl%
\url{https://doi.org/10.1145/1538788.1538814}
\showDOI{\tempurl}


\bibitem[Mcilroy et~al\mbox{.}(2019)]%
        {spectre-here-to-stay}
\bibfield{author}{\bibinfo{person}{Ross Mcilroy}, \bibinfo{person}{Jaroslav
  Sevcik}, \bibinfo{person}{Tobias Tebbi}, \bibinfo{person}{Ben~L. Titzer},
  {and} \bibinfo{person}{Toon Verwaest}.} \bibinfo{year}{2019}\natexlab{}.
\newblock \bibinfo{title}{Spectre is Here to Stay: An Analysis of Side-Channels
  and Speculative Execution}.
\newblock
\newblock
\showeprint[arxiv]{1902.05178}~[cs.PL]


\bibitem[{Microsoft}(2025)]%
        {edge-enhanced}
\bibfield{author}{\bibinfo{person}{{Microsoft}}.}
  \bibinfo{year}{2025}\natexlab{}.
\newblock \bibinfo{title}{Enhance Your Security on the Web with {Microsoft
  Edge}}.
\newblock \bibinfo{howpublished}{Microsoft Support}.
\newblock
\urldef\tempurl%
\url{https://support.microsoft.com/en-us/edge/enhance-your-security-on-the-web-with-microsoft-edge}
\showURL{%
\tempurl}


\bibitem[Myreen(2010)]%
        {myreen-jit}
\bibfield{author}{\bibinfo{person}{Magnus~O. Myreen}.}
  \bibinfo{year}{2010}\natexlab{}.
\newblock \showarticletitle{Verified Just-in-Time Compiler on x86}. In
  \bibinfo{booktitle}{\emph{ACM SIGPLAN-SIGACT Symposium on Principles of
  Programming Languages (POPL)}}. \bibinfo{pages}{107--118}.
\newblock
\urldef\tempurl%
\url{https://doi.org/10.1145/1706299.1706313}
\showDOI{\tempurl}


\bibitem[Myreen and Davis(2011)]%
        {jitawa}
\bibfield{author}{\bibinfo{person}{Magnus~O. Myreen} {and}
  \bibinfo{person}{Jared Davis}.} \bibinfo{year}{2011}\natexlab{}.
\newblock \showarticletitle{A Verified Runtime for a Verified Theorem Prover}.
  In \bibinfo{booktitle}{\emph{Interactive Theorem Proving (ITP)}}
  \emph{(\bibinfo{series}{LNCS}, Vol.~\bibinfo{volume}{6898})}.
  \bibinfo{pages}{265--280}.
\newblock
\urldef\tempurl%
\url{https://doi.org/10.1007/978-3-642-22863-6_20}
\showDOI{\tempurl}


\bibitem[Nelson et~al\mbox{.}(2020)]%
        {jitterbug}
\bibfield{author}{\bibinfo{person}{Luke Nelson}, \bibinfo{person}{Jacob {Van
  Geffen}}, \bibinfo{person}{Emina Torlak}, {and} \bibinfo{person}{Xi Wang}.}
  \bibinfo{year}{2020}\natexlab{}.
\newblock \showarticletitle{Specification and Verification in the Field:
  Applying Formal Methods to {BPF} Just-in-Time Compilers in the {Linux}
  Kernel}. In \bibinfo{booktitle}{\emph{USENIX Symposium on Operating Systems
  Design and Implementation (OSDI)}}. \bibinfo{pages}{41--61}.
\newblock


\bibitem[Norman(2021)]%
        {edgesdsm}
\bibfield{author}{\bibinfo{person}{Johnathan Norman}.}
  \bibinfo{year}{2021}\natexlab{}.
\newblock \bibinfo{title}{Super Duper Secure Mode}.
\newblock \bibinfo{howpublished}{Microsoft Browser Vulnerability Research
  blog}.
\newblock
\urldef\tempurl%
\url{https://microsoftedge.github.io/edgevr/posts/Super-Duper-Secure-Mode/}
\showURL{%
\tempurl}


\bibitem[{RISC-V International}({[n.\,d.]})]%
        {sail-riscv}
\bibfield{author}{\bibinfo{person}{{RISC-V International}}.}
  \bibinfo{year}{[n.\,d.]}\natexlab{}.
\newblock \bibinfo{title}{The {Sail} {RISC-V} Model}.
\newblock \bibinfo{howpublished}{GitHub repository}.
\newblock
\urldef\tempurl%
\url{https://github.com/riscv/sail-riscv}
\showURL{%
\tempurl}


\bibitem[Sammler et~al\mbox{.}(2022)]%
        {islaris}
\bibfield{author}{\bibinfo{person}{Michael Sammler}, \bibinfo{person}{Angus
  Hammond}, \bibinfo{person}{Rodolphe Lepigre}, \bibinfo{person}{Brian
  Campbell}, \bibinfo{person}{Jean Pichon-Pharabod}, \bibinfo{person}{Derek
  Dreyer}, \bibinfo{person}{Deepak Garg}, {and} \bibinfo{person}{Peter
  Sewell}.} \bibinfo{year}{2022}\natexlab{}.
\newblock \showarticletitle{Islaris: Verification of Machine Code Against
  Authoritative {ISA} Semantics}. In \bibinfo{booktitle}{\emph{ACM SIGPLAN
  Conference on Programming Language Design and Implementation (PLDI)}}.
  \bibinfo{pages}{825--840}.
\newblock
\urldef\tempurl%
\url{https://doi.org/10.1145/3519939.3523434}
\showDOI{\tempurl}


\bibitem[Sewell et~al\mbox{.}(2023)]%
        {cakeml-eval}
\bibfield{author}{\bibinfo{person}{Thomas Sewell}, \bibinfo{person}{Magnus~O.
  Myreen}, \bibinfo{person}{Yong~Kiam Tan}, \bibinfo{person}{Ramana Kumar},
  \bibinfo{person}{Alexander Mihajlovic}, \bibinfo{person}{Oskar Abrahamsson},
  {and} \bibinfo{person}{Scott Owens}.} \bibinfo{year}{2023}\natexlab{}.
\newblock \showarticletitle{Cakes That Bake Cakes: Dynamic Computation in
  {CakeML}}.
\newblock \bibinfo{journal}{\emph{Proc.\ ACM Program.\ Lang.}}
  \bibinfo{volume}{7}, \bibinfo{number}{PLDI} (\bibinfo{year}{2023}),
  \bibinfo{pages}{1121--1144}.
\newblock
\urldef\tempurl%
\url{https://doi.org/10.1145/3591266}
\showDOI{\tempurl}


\bibitem[Smith et~al\mbox{.}(2024)]%
        {icarus}
\bibfield{author}{\bibinfo{person}{Naomi Smith}, \bibinfo{person}{Abhishek
  Sharma}, \bibinfo{person}{John Renner}, \bibinfo{person}{David Thien},
  \bibinfo{person}{Fraser Brown}, \bibinfo{person}{Hovav Shacham},
  \bibinfo{person}{Ranjit Jhala}, {and} \bibinfo{person}{Deian Stefan}.}
  \bibinfo{year}{2024}\natexlab{}.
\newblock \showarticletitle{Icarus: Trustworthy Just-In-Time Compilers with
  Symbolic Meta-Execution}. In \bibinfo{booktitle}{\emph{ACM Symposium on
  Operating Systems Principles (SOSP)}}. \bibinfo{pages}{473--487}.
\newblock
\urldef\tempurl%
\url{https://doi.org/10.1145/3694715.3695949}
\showDOI{\tempurl}


\bibitem[Stone(2022)]%
        {p0-2021}
\bibfield{author}{\bibinfo{person}{Maddie Stone}.}
  \bibinfo{year}{2022}\natexlab{}.
\newblock \bibinfo{title}{The More You Know, The More You Know You Don't Know:
  A Year in Review of 0-days Used In-the-Wild in 2021}.
\newblock \bibinfo{howpublished}{Google Project Zero blog}.
\newblock
\urldef\tempurl%
\url{https://projectzero.google/2022/04/the-more-you-know-more-you-know-you.html}
\showURL{%
\tempurl}


\bibitem[Tan et~al\mbox{.}(2019)]%
        {cakeml-backend}
\bibfield{author}{\bibinfo{person}{Yong~Kiam Tan}, \bibinfo{person}{Magnus~O.
  Myreen}, \bibinfo{person}{Ramana Kumar}, \bibinfo{person}{Anthony Fox},
  \bibinfo{person}{Scott Owens}, {and} \bibinfo{person}{Michael Norrish}.}
  \bibinfo{year}{2019}\natexlab{}.
\newblock \showarticletitle{The Verified {CakeML} Compiler Backend}.
\newblock \bibinfo{journal}{\emph{Journal of Functional Programming}}
  \bibinfo{volume}{29} (\bibinfo{year}{2019}), \bibinfo{pages}{e2}.
\newblock
\urldef\tempurl%
\url{https://doi.org/10.1017/S0956796818000229}
\showDOI{\tempurl}


\bibitem[Ueyama(2026)]%
        {sail-x86}
\bibfield{author}{\bibinfo{person}{Rui Ueyama}.}
  \bibinfo{year}{2026}\natexlab{}.
\newblock \bibinfo{title}{sail-x86}.
\newblock
  \bibinfo{howpublished}{\url{https://github.com/shina-lab/sail-x86/tree/1829cb6ecc5f924b0fd920582953a2402357a12a}}.
\newblock


\bibitem[VanHattum et~al\mbox{.}(2024)]%
        {crocus}
\bibfield{author}{\bibinfo{person}{Alexa VanHattum}, \bibinfo{person}{Monica
  Pardeshi}, \bibinfo{person}{Chris Fallin}, \bibinfo{person}{Adrian Sampson},
  {and} \bibinfo{person}{Fraser Brown}.} \bibinfo{year}{2024}\natexlab{}.
\newblock \showarticletitle{Lightweight, Modular Verification for
  {WebAssembly}-to-Native Instruction Selection}. In
  \bibinfo{booktitle}{\emph{ACM International Conference on Architectural
  Support for Programming Languages and Operating Systems (ASPLOS)}}.
\newblock
\urldef\tempurl%
\url{https://doi.org/10.1145/3617232.3624862}
\showDOI{\tempurl}


\bibitem[Vishwanathan et~al\mbox{.}(2023)]%
        {agni}
\bibfield{author}{\bibinfo{person}{Harishankar Vishwanathan},
  \bibinfo{person}{Matan Shachnai}, \bibinfo{person}{Srinivas Narayana}, {and}
  \bibinfo{person}{Santosh Nagarakatte}.} \bibinfo{year}{2023}\natexlab{}.
\newblock \showarticletitle{Verifying the Verifier: {eBPF} Range Analysis
  Verification}. In \bibinfo{booktitle}{\emph{Computer Aided Verification
  (CAV)}} \emph{(\bibinfo{series}{LNCS}, Vol.~\bibinfo{volume}{13966})}.
\newblock
\urldef\tempurl%
\url{https://doi.org/10.1007/978-3-031-37709-9_12}
\showDOI{\tempurl}


\bibitem[Wang et~al\mbox{.}(2014)]%
        {jitk}
\bibfield{author}{\bibinfo{person}{Xi Wang}, \bibinfo{person}{David Lazar},
  \bibinfo{person}{Nickolai Zeldovich}, \bibinfo{person}{Adam Chlipala}, {and}
  \bibinfo{person}{Zachary Tatlock}.} \bibinfo{year}{2014}\natexlab{}.
\newblock \showarticletitle{Jitk: A Trustworthy In-Kernel Interpreter
  Infrastructure}. In \bibinfo{booktitle}{\emph{USENIX Symposium on Operating
  Systems Design and Implementation (OSDI)}}. \bibinfo{pages}{33--47}.
\newblock


\bibitem[yu~Guo and Palsberg(2011)]%
        {traces}
\bibfield{author}{\bibinfo{person}{Shu yu Guo} {and} \bibinfo{person}{Jens
  Palsberg}.} \bibinfo{year}{2011}\natexlab{}.
\newblock \showarticletitle{The Essence of Compiling with Traces}. In
  \bibinfo{booktitle}{\emph{ACM SIGPLAN-SIGACT Symposium on Principles of
  Programming Languages (POPL)}}. \bibinfo{pages}{563--574}.
\newblock
\urldef\tempurl%
\url{https://doi.org/10.1145/1926385.1926450}
\showDOI{\tempurl}


\bibitem[Yuan et~al\mbox{.}(2022)]%
        {certrbpf}
\bibfield{author}{\bibinfo{person}{Shenghao Yuan},
  \bibinfo{person}{Fr{\'e}d{\'e}ric Besson}, \bibinfo{person}{Jean-Pierre
  Talpin}, \bibinfo{person}{Samuel Hym}, \bibinfo{person}{Koen Zandberg}, {and}
  \bibinfo{person}{Emmanuel Baccelli}.} \bibinfo{year}{2022}\natexlab{}.
\newblock \showarticletitle{End-to-End Mechanized Proof of an {eBPF} Virtual
  Machine for Micro-controllers}. In \bibinfo{booktitle}{\emph{Computer Aided
  Verification (CAV)}} \emph{(\bibinfo{series}{LNCS},
  Vol.~\bibinfo{volume}{13372})}.
\newblock
\urldef\tempurl%
\url{https://doi.org/10.1007/978-3-031-13188-2_15}
\showDOI{\tempurl}


\end{thebibliography}

\end{document}